\documentclass[a4paper]{article}
\usepackage{AlexStyle}
\usepackage{import}
\usepackage{color}
\usepackage[colorlinks=true,citecolor=green]{hyperref}
\usepackage{ulem}
\usepackage{ esint }

\numberwithin{equation}{section}

\definecolor{shadecolor}{rgb}{0.95, 0.95, 0.86}
\usetikzlibrary{decorations.markings}

\def\red#1{\textcolor[rgb]{0.9, 0, 0}{#1} }

\def\blue#1{\textcolor[rgb]{0,0,1}{#1}}

\def\be{\begin{equation}}
\def\ee{\end{equation}}
\def\bi{\begin{itemize}}
\def\ei{\end{itemize}}

\def\bea{\begin{eqnarray}}
\def\eea{\end{eqnarray}}
\def\bl{\begin{lemma}}
\def\el{\end{lemma}}
\def\bd{\begin{definition}}
\def\ed{\end{definition}}
\def\bp{\begin{proposition}}
\def\ep{\end{proposition}}

\def\br{\begin{remark}}
\def\er{\end{remark}}

\def\bt{\begin{theorem}}
\def\et{\end{theorem}}
\def\bc{\begin{corollary}}
\def\ec{\end{corollary}}
\renewcommand{\d}{\delta}
\renewcommand{\o}{\omega}
\def\m{\mu}

\def\r{\rho}
\def\tk{\tilde k}
\def\to{\tilde \o}
\def\dn{\rm  dn}

\begin{document}

\title{ Spectral theory of soliton gases for the defocusing NLS equation} 

\author[1]{Alexander Tovbis\footnote{Alexander.Tovbis@ucf.edu}}  
\author[2,3]{Fudong Wang\footnote{fudong@gbu.edu.cn}}

\affil[1]{University of Central Florida, Orlando FL, U.S.A.}
\affil[2]{School of Sciences, Great Bay University, Dongguan, China,
}
\affil[3]{Great Bay Institute For Advanced Study, Dongguan, China }


\maketitle

\begin{abstract} 
In this paper we derive  Nonlinear Dispersion Relations (NDR) for the defocusing NLS    (dark)  soliton gas using the idea of thermodynamic limit  of quasimomentum and quasienergy differentials on  the underlying family  of Riemann surfaces. It turns out that the obtained NDR are closely connected with the recently studied NDR for circular soliton gas for the focusing NLS.  We find solutions for the kinetic equation for  the defocusing NLS  soliton condensate, which is defined by the endpoints of the spectral support $\G$ for the NDR.It turns out that, similarly to KdV soliton condensates (\cite{CERT}), the evolution of these endpoints is governed by the defocusing NLS-Whitham equations (\cite{Kodama}).
We also study the Riemann problem for step initial data
and the kurtosis of genus zero and one defNLS condensates, where we proved that  the kurtosis of genus one condensate can not exceed 3/2, whereas for genus zero condensate the kurtosis is always 1.

\end{abstract}

\tableofcontents

 \section{Introduction}\label{sec-intro}
 
 Spectral theory of soliton gases can be traced back to the paper of V. Zakharov \cite{Za71}, where the effective average velocity of a Korteweg-de Vries (KdV) soliton on the background of many other solitons was first derived.  As it is well known, two such solitons moving with different velocities retain their shapes and velocities after the interaction  but experience a phase shift. That is, the result of this interaction  can be viewed as an ``abrupt" shift of location of the center of each soliton involved.  Naturally, the summary effect of consistent  interactions changes the average velocity of a soliton,
 and that average velocity was derived in  \cite{Za71}.  This property of soliton interactions is, to some extend, common for many integrable equations. It allows one to interpret  ensembles of, say, KdV solitons as particles of some gas with a prescribed law of pairwise interactions. We will call it a  KdV soliton gas.
 
 In terms of the current KdV soliton gas theory, V. Zakharov's formula described the average velocity of a soliton in the diluted KdV soliton gas. Here the word ``diluted" means that solitons are so sparse that it is possible to distinguish individual soliton-soliton interactions or, in other words, that the spatial density of solitons is small. A successful attempt to derive the average velocity in the dense KdV soliton gas, pioneered by G. El in \cite{El2003}, was based on the idea of thermodynamic limit of finite-gap or nonlinear multiphase solutions to KdV.  Multiphase KdV solutions are spectrally characterized by a finite numbers of segments (bands) on the  spectral line $\R$, where each band corresponds  to a particular mode. If one fixes the center   of each band and starts to shrink these bands to their centers,
 one obtains in the limit a finite set of points (centers) on $\R$ that spectrally characterize a multi-soliton solution (each spectral point corresponds to a soliton).
  The wavenumbers  and frequencies of nonlinear $N$-phase solutions can be described through the periods of the quasimomentum and quasienergy meromorphic differentials on the hyperelliptic Riemann surface(RS) $\Rscr_N$, 
  defined by the endpoints of the bands.
   In particular, the branchcuts of $\Rscr_N$ consist of all the bands together with an additional branchcut going to $\infty$.
 The idea of thermodynamic limit is to increase the number $N$ of modes (bands) while simultaneously shrinking the size of each spectral band at a certain special rate. In the limit, the centers of spectral bands will be  distributed at a some compact $\G\subset \R$ with some positive probability density function $\phi(z)$.
 $\G$ is usually assumed to be a finite set of segments sometimes referred to as superbands.
 
A soliton is uniquely defined by its spectral data, that is, by some $z\in \R$ and a real norming   constant $c$. The value of $z$ does not change in the process of   time $t$ evolution  of the soliton  whereas $c$ is multiplied by a factor $e^{8tz^{\frac 32}}$.
At the same time, a nonlinear multiphase solution is defined by its collection of  bands together with a real phase associated with each band. Like in  the soliton case,  the bands do not evolve with $t$ whereas phases undergo $t$ evolution similar to that of the norming constants.  Existence of the  thermodynamic limit of nonlinear multiphase solutions is generically not clear.  However, under the right scaling of the bands (with $N$) one can expect the existence of {\it continualized} limits of wavenumbers and frequencies of these solutions, called density of  states (DOS)  $u(z)$ and density of fluxes (DOF) $v(z)$ respectively. In fact, the DOS and DOF   were  introduced in \cite{El2003} as solutions of some Fredholm type integral equations. These equations form nonlinear dispersion relations (NDR) for the corresponding soliton gas, as they connect the $u$ and $v$ through the underlying 
thermodynamic limit of  meromorphic differentials on RS  $\Rscr_N$. It is interesting to mention that the kernel of the integral operator in the NDR coincides with the phase shift expression for two interacting solitons.

The ratio $s(z)=\frac{v(z)}{u(z)}$ represents the average velocity of the element of the soliton  gas parameterized by $z$. By replacing $v$ by $s$ and combining the two NDR, one can obtain a single integral equation connecting $u$ and $s$, called the {\it equation of states}. This equation (\cite{El2003}) is a direct generalization of Zakharov's average velocity expression from \cite{Za71} to a dense soliton gas.  

So far  we described the {\it homogeneous} KdV soliton gas, that is, the case when $u,v$, as well as the spectral support $\G$ does not depend on $x,t$. Assuming such dependence on very large scales, one comes to the idea of nonhomogeneous soliton gas, in which case the equation of states is complemented by conservation equation $u_t+v_x=0$. These two equations together are called {\it kinetic equation}, see \cite{El2003}.

The NLS equation has the form
\begin{equation}  \label{NLS}
	i  \psi_t +  \psi_{xx} +2 s|\psi|^2 \psi=0,
\end{equation}
where $x,t\in \R$ are the space-time variables, $s=\pm 1$ and  $\psi:\R^2 \ra \C$ is the unknown complex-valued function.  This equation describes the evolution of a slowly varying envelope of a quasi-monochromatic wave packet propagating through a   focusing ($s=1$) or defocusing ($s=-1$)  dispersive  nonlinear  medium.

The  NDR and the kinetic equation for the focusing NLS (fNLS) soliton gases were derived in \cite{ET2020}. 
 The main distinctions from the KdV case is that the fNLS solutions, including soliton solutions, are complex valued and are  spectrally characterized   by Schwarz symmetric sets in the $\C$. For example, a solitary wave (soliton) is spectrally represented by a pair of complex conjugated points $z,\bar z$, where $2|\Im z|$   represent the amplitude of the soliton and $-4\Re z$ represents its velocity.  Therefore, the spectral support set $\G$ is s Schwarz symmetrical set in $\C$, and the integral equations forming the NDR are complex valued. Below, we very briefly  describe the construction of  the NDR for fNLS, starting with a 
genus $N$ Schwarz symmetrical hyperelliptic  Riemann surface $\Rscr_N$.

The  wavenumbers $k_j$ and frequencies $\o_j$ of any finite gap solution to the fNLS, associated with  $\Rscr_N$, are defined in terms of  $\Rscr_N$.  In particular, let $dp_N$, $dq_N$ be the second kind real normalized meromorphic differentials on  $\Rscr_N$ with poles only at infinity (both sheets) and with the principle  parts $\pm 1$ for $dp_N$   and  $\pm 2z$ for $dq_N$ there respectively. Here $z\in \Rscr_N$ denotes the spectral parameter.
and  real normalized mean that all all the periods of $dp_N$, $dq_N$ on $\Rscr_N$ are real.
Then the vectors $\vec k$,   $\vec \o$ of the (real) periods of   $dp_N$, $dq_N$ with respect to a fixed homology basis ($\bf A$ and $\bf B$ cycles) of $\Rscr_N$ are vectors of the wavenumbers and frequencies of finite gap solutions on  $\Rscr_N$ respectively, i.e., 
\be\label{waven-freq}
k_j=\oint_{\bf A_j}dp_N,~~~\tk_j= \oint_{\bf B_j}dp_N,\quad
\o_j=\oint_{\bf A_j}dq_N,~~~\to _j= \oint_{\bf B_j}dq_N,
\ee
where $\vec k=(k_1,\dots,k_N,\tk_1,\dots,\tk_N),$ $\vec \o=(\o_1,\dots,\o_N,\to_1,\dots,\to_N)$.
The differentials  $dp_N$, $dq_N$  are known as quasimomentum and quasienergy differentials respectively (
\cite{ForLee}).

Denote by $w_{j,N}=w_j$ the j-th normalized holomorphic differential on $\Rscr_N$, $j=1,\dots,N$, that are defined by the condition $\int_{A_k}w_j =\d_{k,j}$, $k=1,\dots,N$, where $\d_{k,j}$ is the Kronecker delta. The well known Riemann Bilinear Relations (RBR)

\bea
\sum_j\left[\oint_{\bf A_j}w_m\oint_{\bf B_j}dp_N-\oint_{\bf A_j}dp_N\oint_{\bf B_j}w_m\right]=2\pi i {\sum} {\rm Res}(\int w_m dp_N), \label{RBR1}\\
\sum_j\left[\oint_{\bf A_j}w_m\oint_{\bf B_j}dq_N-\oint_{\bf A_j}dq_N\oint_{\bf B_j}w_m\right]=2\pi i {\sum} {\rm Res}(\int w_m dq_N),   
\label{RBR2}
\eea

$m=1,\dots,N$, form systems of linear equations for $\vec k$,   $\vec \o$  respectively, where the summation in the right hand side is taking over the only poles $z=\infty_\pm$ (infinities on both sheets of $\Rscr_N$) of $dp_N,dq_N$.  
Taking imaginary and  real  parts of \eqref{RBR1}, one gets
\bea
\sum_j k_j \Im \oint_{\bf B_j}w_m=-2\pi \Re\le(  \sum {\rm Res}(\int w_m dp_N)   \ri), \label{RBR1-Im} \\
\tk_m  =  \sum_j k_j \Re \oint_{\bf B_j}w_m-2\pi \Im\le(  \sum {\rm Res}(\int w_m dp_N)   \ri),
\label{RBR1-Re}
\eea
$m=1,\dots,N$, respectively for  \eqref{RBR1}.  Similarly, we can calculate the imaginary and real parts of \eqref{RBR2}.  We observe that
the matrix of the $N\times N$ system of linear equations   \eqref{RBR1-Im}
 is positive definite, since it is the imaginary part $\Im \tau$ of the Riemann period matrix $\tau= \oint_{\bf B_j}w_m$.  Once $k_j$ are known, the values of $\tk_j$ can be calculated from \eqref{RBR1-Re}.
Thus, the systems  \eqref{RBR1-Im}- \eqref{RBR1-Re} always have a unique solution.  
Similar results are true for \eqref{RBR2}.
As the wavenumber and frequency vectors $\vec k$, $\vec \o$ are connected via the  underlying Riemann surface $\Rscr_N$, we  can interpret 
the RBR   \eqref{RBR1}- \eqref{RBR2}  as a (discrete)
 NDR for the finite gap solutions to the fNLS that are  defined by  $\Rscr_N$. 

One of the  main subjects of the spectral theory of soliton gases is the thermodynamic limit of scaled vectors  $\vec k$, $\vec\o$, i.e., the thermodynamic limit  of the RBR \eqref{RBR1}-\eqref{RBR2}, which leads to a continualized version of the NDR established in \cite{ET2020}, see equations 
	 \eqref{NDR}-\eqref{NDR2} and \eqref{dr_breather_gas1}-\eqref{dr_breather_gas2} below.

The  connection between a finite gap and a soliton solution can be illustrated on the example of  an elliptic solution  $\psi_m(x,t) = e^{it(2-m)}{\dn}(x,m)$ to the fNLS \eqref{NLS}, where $\dn$ denotes the corresponding Jacobi elliptic function and $m\in[0,1]$ is the elliptic parameter.  Spectrally, this solution is represented by two symmetric bands $\G^\pm \subset[-i,i]$, whose endpoints depend on $m$. In fact 
$\G^+=\frac i2[1-\sqrt{1-m},1+\sqrt{1-m}]$, where $\sqrt{1-m},1$ are the minimum and the maximum values of $\dn(x,m) $ on $x\in\R$ respectively. In the limit $m=0$ we obtain the plane wave    $\psi_0(x,t) = e^{2it}$ with
the spectral support $\G^+=[0,i]$
In the  opposite limit we obtain the soliton solution
 $\psi_1(x,t) = e^{it}\sech(x) $
 with the spectral support $\G^+=\{\hf i\}$.
  That is, the period of an elliptic solution tends to infinity if one  shrinks the spectral bands $\G^\pm$ of this solution. Clearly, the corresponding wavenumber tends to zero.

For simplicity of describing the thermodynamic limit, let us assume that $N$ is even and only one (Schwarz symmetrical) band $\g_0$ intersects $\R$.  We choose 
the $\bf A$ cycles as closed loops around every band except $\g_0$. Then the $\bf B$ cycle  
$\bf B_j$ is 
represented by  a  properly oriented loop going through  the branch cut encircled by $\bf A_j$ and through
the exceptional branch cut  $\g_0$, $j=1,\dots,N$.

In the thermodynamic limit, we assume 
(see \cite{ET2020})
	 that the number $N+1$ of  bands (the branch cuts of $\Rscr_N$) is growing so that in the limit the centers of  bands are distributed on some (Schwarz symmetrical) compact $\G$ in $\C$ with a 
	normalized  continuous (Schwarz symmetrical) density $\phi(z)>0$, that is,  $\int_{\G^+}\phi(w)|dw|=1$, where  on $\G^+=\G\cap\C^+$. 

Simultaneously, the bandwidth of each band (except, possibly, $\g_0$) centered at $z\in\G^+$   is shrinking at the order $e^{-\nu(z)N}$, where $\nu(z)$ is a continuous non-negative function on $\G^+$,
in such a way that the distance between any two bands should be of the order at least $O(1/N)$.  The wavenumbers $k_j$ and frequencies $\o_j$ are called solitonic wavenumbers and frequencies respectively, because in the thermodynamic limit they go to zero. 
The remaining quantities $\tilde k_j, \tilde \o_j$ from \eqref{waven-freq}  are called carrier wavenumbers and frequencies respectively.
The function $\s(z)=\frac{ 2\nu(z)}{\phi(z)}$ is called spectral scaling function. 
	In 
	the thermodynamic limit, the system of linear equations \eqref{RBR1-Im}  for the solitonic wavenumbers $k_j$, subject to additional assumption that the size of the exceptional band $\g_0$ is also shrinking, turns into the integral equation  \eqref{NDR} for the scaled continualized limit $u(z)$ of  $k_j$:
	\begin{align}\label{NDR}
		\int_{\G^+}\log\le|\frac{z-\bar w}{z-w}\ri|u(w)|dw|+\sigma(z)u(z)&=\Im{z},\quad z\in \G^+,\\
		\int_{\G^+}\log\le|\frac{z-\bar w}{z-w}\ri|v(w)||dw|+\sigma(z)v(z)&=-4\Re{z}\Im{z},\quad z\in \G^+.
		\label{NDR2}
	\end{align}
	Similarly, the imaginary part of \eqref{RBR2} for the solitonic frequencies  $\o_j$ turns into the integral equation  \eqref{NDR2} for the scaled continualized limit $v(z)$ of  $\o_j$.
	Thus, equations
	\eqref{NDR}-\eqref{NDR2}
	represent the thermodynamic limit  of the systems of linear equations for $k_j,\o_j$, i.e.,  the thermodynamic limit of imaginary parts of the  RBR \eqref{RBR1}-\eqref{RBR2}. They form the  NDR for fNLS soliton gas.
	If we require that the exceptional band $\g_0$ stays unchanged, say,  $\g_0=[-\d_0,\d_0]$, $\d_0\in i\R^+$,
	the thermodynamic limit  will correspond to the fNLS breather gas. 
	 The 	NDR for the solitonic component of the fNLS breather  gas have the form (\cite{ET2020})
	\begin{multline} \label{dr_breather_gas1}
		\int_{\G^+}
		\le[\log\le| \frac{w-\bar z}{w-z}\ri|+ \log\le|\frac{R_0(z)R_0(w)+z w -\d_0^2}
		{R_0(\bar z)R_0(w)+\bar z w-\d_0^2}\ri|\ri]   u(w)|dw| 
		+\sigma(z)u(z) \\
		= \Im R_0(z), 
	\end{multline}
	\begin{multline}  \label{dr_breather_gas2}
		\int_{\Gamma^+}
		\left[\log\left| \frac{w-\bar z}{w-z}\right|+\log\left|\frac{R_0(z)R_0(w)+ z w-\delta_0^2}
		{R_0(\bar z)R_0(w)+ \bar z w-\delta_0^2}\right|\right] v(w) |dw|
		+\sigma(z)v(z) \\
		= - 2\Im[z R_0(z)],
	\end{multline}
	where  	$z\in\Gamma^+$ and $R_0(z)=\sqrt{z^2-\delta_0^2}$ with the branchcut $\g_0=[-\d_0,\d_0]\subset i\R$. 
	
	\br\label{rem-ph-shifts}
	An interesting observation  for both KdV (\cite{El2003}) and fNLS  (\cite{ET2020}) soliton gases is that the kernel of the integral operator of the 1st NDR divided by its the right hand side, i.e.,
	$ \frac 1 {\Im z}\log\le|\frac{z-\bar w}{z-w}\ri|$ for equation \eqref{NDR},
represents the  well known phase shift formula of two interacting solitons with the corresponding spectral parameters $z,w$, see \cite{ZS}.  In fact,  this observation can be extended to phase shifts  of interacting breathers and the 1st NDR \eqref{dr_breather_gas1} for fNLS breather gases(\cite{ET2020}, \cite{Breather}).  In Remark \ref{rem-phase-shift}, Section \ref{sec-derivation} we  expand this observation to the defocusing NLS
(defNLS) soliton gases and  phase shifts  of interacting dark solitons. 
\er

In the present paper we derive the NDR (Section \ref{sec-derivation}) and the kinetic equation (Section \ref{sec-kinetic}) for the defNLS soliton gases, where the solitons are dark (depression) solitons on the plane wave background.
The kernel of the  integral operator in the obtained NDR, see \eqref{1stNDR}-\eqref{2ndNDR},
resemble that of the NDR \eqref{dr_breather_gas1}- \eqref{dr_breather_gas2} for the fNLS breather gas, if we anti Schwarz symmetrically extend  $u(z), v(z)$ there to the lower half plane and integrate over $\G=\G^+\cup\overline{\G^+}$.
 In Section \ref{sec-solution} we show that the NDR \eqref{1stNDR}-\eqref{2ndNDR} for the defNLS soliton gas can be transformed into the NDR for the circular fNLS soliton gas, which was studied in recent paper  \cite{TW2023} of the authors.
We remind that an fNLS soliton gas is called circular if $\G^+$ is restricted to the upper semicircle  $|z|=\r$ for some $\r>0$. Existence and uniqueness of solutions $u(z), v(z)$ to  the NDR \eqref{1stNDR}-\eqref{2ndNDR} follow then from the corresponding results for fNLS soliton gases.

In the case the spectral scaling function  $\s(z)\equiv 0$ on $\G^+$, i.e., in  the case of sub-exponential decay of the bandwidths in the thermodynamic limit, a soliton gas is  called soliton condensate (\cite{ET2020}). fNLS soliton condensates have a natural extremal property: if   $\G^+$ in \eqref{NDR} is fixed  but $\s(z)\geq 0,~\s\in C(\G^+)$, is allowed to vary, the maximal average intensity of the fNLS soliton gas is attained at $\s(z)\equiv 0$. That is,  fNLS soliton condensates maximize the average intensity of fNLS soliton gases with a given $\G^+$, see \cite{KT2021}, i.e., they represent ``most dense" soliton gases. Since defNLS dark solitons represent localized depressions, one could expect that defNLS soliton condensates represent soliton gases of ``least" average density. Indeed, in Section \ref{sec-quasi} (see \eqref{vac}) we show that the defNLS soliton condensate whose spectral support $\G$ coincides with that of the background plane wave is vacuous.

Existence and uniqueness of solutions $u(z),v(z)$ for fNLS soliton gases (under some mild assumptions) was established in  \cite{KT2021}, where it was shown that $u(z)dz$, $v(z)dz$ are supported on $\G^+$ minimizing measures for the Green energy (in $\C^+$)  with external fields $-2\Im z$ and $4\Im z^2$ respectively.  In some special  cases these measures, i.e. the DOS $u(z)$ and DOF $v(z)$ (the latter corresponds to  a signed measure),  can be calculated explicitly. That is the case, for example, for circular fNLS condensates (\cite{TW2023}). In Section \ref{sec-condens} we use the results of (\cite{TW2023}) to construct
solutions $u(z),v(z)$ for defNLS soliton condensates and, in particular, to describe in details solutions of genus zero and genus 1 condensates.  We also showed that $udz,vdz$ are proportional to the quasimomentum and quasienergy differentials on the ``superband" hyperelliptic Riemann surface $\Rscr$ defined by the spectral support $\G$, see Remark \ref{rem-dp-infty}.
In Section \ref{sec-kinetic} we show that, similarly to the KdV soliton condensates, the modulation equations for the endpoints of the spectral support $\G$ of the NDR coincide with the defNLS-Whitham  equations for defNLS slowly modulated finite gap solutions.  We also describe the dynamics of nonhomogeneous genus zero and one soliton condensates with discontinuous  initial conditions.

 In Sections \ref{sec-quasi}
we calculate the thermodynamic limit  of quasimomentum and quasienergy differentials and 
average  densities and fluxes of soliton condensates, which are expressed through the coefficients of Laurent expansions at $z=\infty$ of these differentials.
We show that these limits coincide with  quasimomentum and quasienergy differentials on the superband RS $\Rscr$, see Theorem \ref{th-limdp}. Thus, the average densities and fluxes of defNLS soliton condensates coincide with the corresponding densities and fluxes of the finite gap defNLS solutions defined by $\Rscr$.
 The results of Sections \ref{sec-quasi} are used in Section \ref{sec-kurt} to calculate the average kurtosis $\kappa$ of the realizations of defNLS soliton condensates. We proved that the average kurtosis for genus one defNLS soliton condensates varies between $1$ and $3/2$. As expected, this is in contrast with fNLS condensates, where $\kappa\geq 2$, which corresponds to normal ($\kappa=2$) or fat-tail ($\kappa>2$) distribution of intensity. We also found that, similarly to KdV (\cite{CERT}),
 realizations of genus zero soliton condensate almost surely coincide with a corresponding genus zero defNLS  solution. 
 In Section \ref{sec-diluted} we consider a diluted defNLS condensate and show that its kurtosis varies between $1$ and $2$.

 \paragraph*{Acknowledgement.} 
 The authors would like to thank the Isaac Newton Institute for Mathematical Sciences (INI), Cambridge, and the University of Northumbria, Newcastle, for support and hospitality during the programme "Emergent phenomena in nonlinear dispersive waves" in Summer 2024, where  part of the work on this paper was undertaken. AT  was partially supported by the Simons Foundation Fellowship during this INI programme.
 The work of AT was supported in part by NSF grants  DMS-2009647 and  DMS-2407080. F.Wang is supported by Guangdong Basic and Applied Basic Research Foundation B24030004J.

  \section{NDR for the defocusing NLS soliton gas}\label{sec-derivation}
  
  As it is well known, the spectrum of defocusing NLS (defNLS) belongs to $\R$ (the corresponding ZS problem is self-adjoint). Assume that background  wave is spectrally represented by the segment $I=[-\d_0,\d_0]$,  see, for example, \cite{BBEIM}, where, for simplicity, we will assume $\d_0=1$.  It is well known that nontrivial defNLS solitons are only solitons on nonzero background,
  known as gray (dark) solitons. 
  As it is discussed below, they correspond to spectral parameter $z\in I$.
  So, considering soliton gas for defNLS, we consider the following two spectral settings:
  \begin{itemize}
\item A) all $z_j\in I$;
\item B) all $z_j\not\in I$.
  \end{itemize}
 As we will discuss below,
 only zero solitonic wavenumbers $k_m$ correspond for the option B), whereas the RBR  provide nontrivial  $k_m$ for the option  A). 
 
 We start with considering the RBR for the case B), which is very similar to  the case of a bound state breather  gas for fNLS with the following caveats: i) there is no Schwarz symmetry, i.e., like in the KdV case, a dark soliton  is represented by  a single point $z_j\in\R$, not a pair of symmetrical points; ii) $\G\subset \R\setminus I$. Following \cite{ET2020}, \cite{TW22}, we introduce a hyperelliptic RS $\Rscr_N$ of genus $N$, where all $N$ bands (branchcuts) $\g_j\subset \R\setminus I$, $j=1,\dots,N$, are shrinking much faster than $\frac 1N$, for example,
 exponentially fast in $N$.  
 (We assume  the remaining band $\g_0:=I$ of $\Rscr_N$ to be stationary.) 
 Then, the band normalized holomorphic differentials $w_j$, $j=1,\dots,N$, on  $\Rscr_N$  can be approximated by  
 \be\label{appr-omega_B} 
 w_j\approx -\frac{R_0(z_j)dz}{2\pi iR_0(z)(z-z_j)}
 \ee
 for sufficiently large $N$, where $R_0=\sqrt{z^2-1}$. Assuming the ${\bf A}$ cycles are small negatively oriented bands around each $\g_j$, $j=1,\dots,N$, one can check that $\oint_{{\bf A}_k}w_j=\d_{k,j}$, where $\d_{k,j}$ denotes Kronecker delta. 
  The RBR  can be written as
 \be\label{RBI-B}
 \sum_m\le[ \oint_{{\bf A}_m}w_j  \oint_{{\bf B}_m}dp_N -  \oint_{{\bf B}_m}w_j    \oint_{{\bf A}_m}dp_N \ri]=-4\pi i     \Res (p_N w_j)|_{z=\infty_+}, \quad j=1,\dots,N,
 \ee
 where each ${\bf B}_m$ is a properly oriented loop on both sheets of $\Rscr_N$ passing through $I$ and $\g_m$ 
 and 
 \be\label{dpN}
 dp_N=\frac{P_N(z)dz}{R_0(z)R(z)}
 \ee
  is the real normalized quasimomentum differential.  Here $P_N(z)$ is a monic polynomial of degree $N+1$ with zero residue and
  \be\label{R}
  R(z)=\sqrt{ \prod_{j=1}^{N} (z-\a_{2j-1})   (z-\a_{2j}) },
  \ee
  where $\g_j=[\a_{2j-1},\a_{2j}]$. The real normalization of $dp_N=f(z)dz$ requires that $f(z)$ is Schwarz symmetric, i.e., that all the coefficients of the polynomial $P_N$ are real. Indeed, otherwise, $dp_N$ would be not unique, as $\overline{f(\bar z)}dz$ would be another real normalized 
  meromorphic differential with the same principle parts at $z=\infty$ on both sheets. Since $f(z)$ is purely imaginary on the bands, the real normalization implies that
    all the solitonic wavenumbers $k_m= \oint_{{\bf A}_m}dp_N$, see \eqref{waven-freq}, must be zero.
Then \eqref{RBI-B} imply that the carrier wavenumbers  $\tilde k_m= \oint_{{\bf B}_m}dp_N=-2R_0(z_m)$.
  
\begin{figure}[ht]
	\centering
	\resizebox{0.9\textwidth}{!}{
		\tikzset{every picture/.style={line width=0.75pt}} 

\begin{tikzpicture}[x=0.75pt,y=0.75pt,yscale=-1,xscale=1]
	
	\draw    (340.19,170.67) -- (395.52,170.67) ;
	\draw [shift={(395.52,170.67)}, rotate = 0] [color={rgb, 255:red, 0; green, 0; blue, 0 }  ][fill={rgb, 255:red, 0; green, 0; blue, 0 }  ][line width=0.75]      (0, 0) circle [x radius= 2.01, y radius= 2.01]   ;
	\draw [shift={(340.19,170.67)}, rotate = 0] [color={rgb, 255:red, 0; green, 0; blue, 0 }  ][fill={rgb, 255:red, 0; green, 0; blue, 0 }  ][line width=0.75]      (0, 0) circle [x radius= 2.01, y radius= 2.01]   ;
	\draw    (100.02,171.04) -- (149.99,170.79) ;
	\draw [shift={(149.99,170.79)}, rotate = 359.71] [color={rgb, 255:red, 0; green, 0; blue, 0 }  ][fill={rgb, 255:red, 0; green, 0; blue, 0 }  ][line width=0.75]      (0, 0) circle [x radius= 2.01, y radius= 2.01]   ;
	\draw [shift={(100.02,171.04)}, rotate = 359.71] [color={rgb, 255:red, 0; green, 0; blue, 0 }  ][fill={rgb, 255:red, 0; green, 0; blue, 0 }  ][line width=0.75]      (0, 0) circle [x radius= 2.01, y radius= 2.01]   ;
	\draw    (447.33,169.92) -- (502.67,169.92) ;
	\draw [shift={(502.67,169.92)}, rotate = 0] [color={rgb, 255:red, 0; green, 0; blue, 0 }  ][fill={rgb, 255:red, 0; green, 0; blue, 0 }  ][line width=0.75]      (0, 0) circle [x radius= 2.01, y radius= 2.01]   ;
	\draw [shift={(447.33,169.92)}, rotate = 0] [color={rgb, 255:red, 0; green, 0; blue, 0 }  ][fill={rgb, 255:red, 0; green, 0; blue, 0 }  ][line width=0.75]      (0, 0) circle [x radius= 2.01, y radius= 2.01]   ;
	\draw  [dash pattern={on 0.84pt off 2.51pt}]  (412.19,169.83) -- (436,169.83) ;
	\draw    (270.02,171) -- (320,170.75) ;
	\draw [shift={(320,170.75)}, rotate = 359.71] [color={rgb, 255:red, 0; green, 0; blue, 0 }  ][fill={rgb, 255:red, 0; green, 0; blue, 0 }  ][line width=0.75]      (0, 0) circle [x radius= 2.01, y radius= 2.01]   ;
	\draw [shift={(270.02,171)}, rotate = 359.71] [color={rgb, 255:red, 0; green, 0; blue, 0 }  ][fill={rgb, 255:red, 0; green, 0; blue, 0 }  ][line width=0.75]      (0, 0) circle [x radius= 2.01, y radius= 2.01]   ;
	\draw  [dash pattern={on 0.84pt off 2.51pt}]  (232.19,170.83) -- (256.86,170.83) ;
	\draw    (175.52,171) -- (217,170.92) ;
	\draw [shift={(217,170.92)}, rotate = 359.88] [color={rgb, 255:red, 0; green, 0; blue, 0 }  ][fill={rgb, 255:red, 0; green, 0; blue, 0 }  ][line width=0.75]      (0, 0) circle [x radius= 2.01, y radius= 2.01]   ;
	\draw [shift={(175.52,171)}, rotate = 359.88] [color={rgb, 255:red, 0; green, 0; blue, 0 }  ][fill={rgb, 255:red, 0; green, 0; blue, 0 }  ][line width=0.75]      (0, 0) circle [x radius= 2.01, y radius= 2.01]   ;
	\draw  [color={rgb, 255:red, 74; green, 144; blue, 226 }  ,draw opacity=1 ] (329.4,171) .. controls (329.4,149.46) and (370.05,132) .. (420.2,132) .. controls (470.35,132) and (511,149.46) .. (511,171) .. controls (511,192.54) and (470.35,210) .. (420.2,210) .. controls (370.05,210) and (329.4,192.54) .. (329.4,171) -- cycle ;
	\draw  [color={rgb, 255:red, 74; green, 144; blue, 226 }  ,draw opacity=1 ] (160,170.87) .. controls (160,143.28) and (242.04,120.92) .. (343.25,120.92) .. controls (444.46,120.92) and (526.5,143.28) .. (526.5,170.87) .. controls (526.5,198.46) and (444.46,220.83) .. (343.25,220.83) .. controls (242.04,220.83) and (160,198.46) .. (160,170.87) -- cycle ;
	\draw [color={rgb, 255:red, 208; green, 2; blue, 27 }  ,draw opacity=1 ]   (138.5,171) .. controls (138,156.25) and (196,157.75) .. (196.26,170.96) ;
	\draw [color={rgb, 255:red, 208; green, 2; blue, 27 }  ,draw opacity=1 ]   (304.5,170.5) .. controls (304,155.75) and (362,157.25) .. (362.26,170.46) ;
	\draw [color={rgb, 255:red, 208; green, 2; blue, 27 }  ,draw opacity=1 ] [dash pattern={on 4.5pt off 4.5pt}]  (138.5,171) .. controls (138,186.25) and (196,186.25) .. (196.26,170.96) ;
	\draw [color={rgb, 255:red, 208; green, 2; blue, 27 }  ,draw opacity=1 ] [dash pattern={on 4.5pt off 4.5pt}]  (304.5,170.5) .. controls (304,185.75) and (362,185.75) .. (362.26,170.46) ;
	\draw [color={rgb, 255:red, 74; green, 144; blue, 226 }  ,draw opacity=1 ]   (309,121.25) .. controls (320.83,120.8) and (330.59,120.75) .. (340.35,120.88) ;
	\draw [shift={(343.25,120.92)}, rotate = 180.89] [fill={rgb, 255:red, 74; green, 144; blue, 226 }  ,fill opacity=1 ][line width=0.08]  [draw opacity=0] (10.72,-5.15) -- (0,0) -- (10.72,5.15) -- (7.12,0) -- cycle    ;
	\draw [color={rgb, 255:red, 208; green, 2; blue, 27 }  ,draw opacity=1 ]   (155,161.25) .. controls (159.38,160.38) and (166.81,160.27) .. (173.63,160.92) ;
	\draw [shift={(176.5,161.25)}, rotate = 187.59] [fill={rgb, 255:red, 208; green, 2; blue, 27 }  ,fill opacity=1 ][line width=0.08]  [draw opacity=0] (10.72,-5.15) -- (0,0) -- (10.72,5.15) -- (7.12,0) -- cycle    ;
	\draw [color={rgb, 255:red, 74; green, 144; blue, 226 }  ,draw opacity=1 ]   (400.44,132.75) .. controls (410.17,132.31) and (418.23,132.26) .. (426.26,132.37) ;
	\draw [shift={(429.24,132.42)}, rotate = 181.06] [fill={rgb, 255:red, 74; green, 144; blue, 226 }  ,fill opacity=1 ][line width=0.08]  [draw opacity=0] (10.72,-5.15) -- (0,0) -- (10.72,5.15) -- (7.12,0) -- cycle    ;
	\draw [color={rgb, 255:red, 208; green, 2; blue, 27 }  ,draw opacity=1 ]   (326.78,160.22) .. controls (332,160.03) and (337.72,160.19) .. (344.06,160.9) ;
	\draw [shift={(346.94,161.25)}, rotate = 187.59] [fill={rgb, 255:red, 208; green, 2; blue, 27 }  ,fill opacity=1 ][line width=0.08]  [draw opacity=0] (10.72,-5.15) -- (0,0) -- (10.72,5.15) -- (7.12,0) -- cycle    ;
	
	\draw (99.64,174.97) node [anchor=north west][inner sep=0.75pt]  [font=\small,rotate=-35.94]  {$a_{0} =-1$};
	\draw (154.55,174.11) node [anchor=north east] [inner sep=0.75pt]  [font=\small]  {$a_{1}$};
	\draw (413.72,174.17) node [anchor=north east] [inner sep=0.75pt]  [font=\small]  {$a_{2j+1}$};
	\draw (554.51,203.03) node [anchor=north east] [inner sep=0.75pt]  [font=\small,rotate=-32.55]  {$a_{2N+1} =1$};
	\draw (342.19,174.07) node [anchor=north west][inner sep=0.75pt]  [font=\small]  {$a_{2j}$};
	\draw (449.33,173.32) node [anchor=north west][inner sep=0.75pt]  [font=\small]  {$a_{2N}$};
	\draw (258.86,174.23) node [anchor=north west][inner sep=0.75pt]  [font=\small]  {$a_{2j-2}$};
	\draw (330.06,175.7) node [anchor=north east] [inner sep=0.75pt]  [font=\small]  {$a_{2j-1}$};
	\draw (174.12,173.5) node [anchor=north west][inner sep=0.75pt]  [font=\small]  {$a_{2}$};
	\draw (227.89,173.4) node [anchor=north east] [inner sep=0.75pt]  [font=\small]  {$a_{3}$};
	\draw (193.6,202.8) node [anchor=north west][inner sep=0.75pt]  [color={rgb, 255:red, 74; green, 144; blue, 226 }  ,opacity=1 ]  {$B_{1}$};
	\draw (335.4,191.4) node [anchor=north west][inner sep=0.75pt]  [color={rgb, 255:red, 74; green, 144; blue, 226 }  ,opacity=1 ]  {$B_{j}$};
	\draw (125.8,146.4) node [anchor=north west][inner sep=0.75pt]  [color={rgb, 255:red, 208; green, 2; blue, 27 }  ,opacity=1 ]  {$A_{1}$};
	\draw (296.6,145.2) node [anchor=north west][inner sep=0.75pt]  [color={rgb, 255:red, 208; green, 2; blue, 27 }  ,opacity=1 ]  {$A_{j}$};

\end{tikzpicture}
}

\caption{The hyperelliptic Riemann surface $\mathcal{R}_N$, indicating the orientation of the homology basis. {Here $[\a_{2j-1},\a_{2j}]$ are the gaps  $c_j$.  The last band is  $[\a_{2N},\a_{2N+1}]$.}}
\label{Fig:Cont}
\vspace{5pt}
\end{figure}
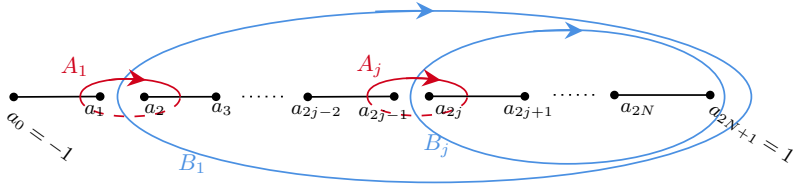 
 Let us turn now to the case A), where $z_j\in I$. Interpreting now each $z_j$ as  a center of a small gap $c_j$, consider
 an $N$-gap defNLS solution defined by the RS $\Rscr_N$ with the gaps $c_j=[\a_{2j-1}, \a_{2j}]$, $j=1,\dots,N$.
 It is easy to check that gap normalized holomorphic differentials are approximated by
 \be\label{appr-omega_A} 
 w_j\approx -\frac{R_0(z_j)dz}{ 2 \pi iR_0(z)\sqrt{(z-z_j)^2-\d_j^2}},
 \ee
 where $\d_j$  denotes  the half-length of the $j$th gap and
 all the gaps $c_j$
 are shrinking much faster than $\frac 1N$.
 Indeed,  choosing ${\bf A}_k$ to be a negatively oriented cycle around   the $k$th gap (twice the integral over the gap), we obtain
$\oint_{{\bf A}_k}w_j=\d_{k,j}$.
The corresponding ${\bf B_j}$ cycle is a negatively oriented loop on the main sheet that contains all but the first $j$ bands (bands and gaps are counted left to right), see Figure \ref{Fig:Cont}. Now the real normalization of $dp$ on  RS $\Rscr_N$ means  that all the carrier wavenumbers $\tilde k_j=0$, whereas the solitonic wavenumbers satisfy
\be\label{RBI-Im}
\Im \tau \vec k= 4\pi (..., \Res (p_N w_j)|_{z=\infty_+}, ...)^T,
\ee
where $\vec k=(k_1,..., k_N)^T$ with  $k_j=\oint_{{\bf A}_j}dp_N$ is the vector of solitonic  wavenumbers, $\tau=(  \oint_{{\bf B}_m}w_j )_{m,j}$ is the Riemann period matrix of 
 $\Rscr_N$ and $T$ denotes the transposition.

 Equation \eqref{RBI-Im} is the 1st discrete NDR for defNLS soliton gas.

We now use the identity
\be\label{anti-der}
	\int\frac{d\z}{R_0(\z)(\z-\eta)}=-\frac{1}{R_0(\eta)}\ln\frac{R_0(\eta)R_0(\z)+\z\eta-1}{\eta-\z}~~~
	\ee
 to calculate the thermodynamic limit of the off-diagonal entries of $\tau$. The asymptotics of diagonal entries is as in  \cite{TW22}, \cite{ET2020}.
Indeed, the cycle ${\bf B}_j$ is the negatively oriented loop around the bands $j+1, ...,N+1$ on the main sheet, i.e.,
 \bea\label{ker-calc}
  \oint_{{\bf B}_m}w_j
   =2\fint_{z_m}^1 w_j \approx - 2 \frac{R_0(z_j)}{ 2\pi i}\fint_{z_m}^1 \frac{dz}{R_0(z)(z-z_j)}=\cr
  \le.\frac{1}{\pi i}\ln\frac{R_0(z_j)R_0(z)+zz_j-1}{-|z_j-z|}\ri|_{z_m}^1 = 
  \frac{1}{\pi i}\ln\frac{-|z_j-z_m|}{R_0(z_j)R_0(z_m)+z_mz_j-1},
 \eea
where we used the fact that the latter expression is purely imaginary (since so is $w_j$ on the bands) and the denominator there is negative. 
 We also approximated $\sqrt{(z-z_j)^2-\d_j^2}$ by $z-z_j$ when integrated over ${\bf B}_m$.  Then the RBR  \eqref{RBI-B}  can be written as
\be\label{RBI-full}  
-\frac i{2\pi}\sum_{m=1}^N\le[  \ln\frac{R_0(z_j)R_0(z_m)+z_mz_j-1}{-|z_j-z_m|}    \ri] k_m\approx -R_0(z_j).
 \ee 
Following the same steps as in \cite{ET2020} in  calculating the thermodynamic limit of \eqref{RBI-full}, we  obtain
 \be\label{1stNDR-full}
 -i\int_\G\le[ \ln\frac{R_0(z)R_0(w)+wz-1}{-|z-w|}
 \ri]u(w)dw-i\s(z)u(z)= -R_0(z),\quad z\in\G,
 \ee
 where $\G\subseteq I$ is the accumulation set of the centers of gaps $z_j$, $j=1,\dots,N$, (often called spectral support set) and 
  functions $u,\s$ 
 are defined exactly as in \cite{ET2020}, see also Section \ref{sec-intro}. Namely, the DOS $u(z)=\frac{\kappa(z)\phi(z)}{2\pi}$ and the spectral scaling function $\s(z)=\frac{2\nu(z)}{\phi(z)}$, where
 	$\phi(z)$ is the probability density for the distribution of the centers of gaps $z_j$,  $\nu(z)\geq 0$ defines the length $2\d_j$ of the gap centered at $z_j$ by $\d_j=e^{-N\nu(z_j)}$   and $\kappa(z)$ is some smooth interpolation of $Nk_j$ in the thermodynamic limit.  As it was mentioned in Section \ref{sec-intro}, the error estimate for the derivation \eqref{1stNDR-full} in the case $\s(z)>0$ was discussed in \cite{TW22}.

 The second RBR with $dp$ replaced by  $dq$ in \eqref{RBI-B} has the same form, with $u$ replaced by $v$ and the RHS  $2zR_0(z)$  instead of $R_0(z)$.
 Thus,
  we obtain  the  NDR for the defNLS (dark) soliton gas:
 \begin{align}
  \int_\G \ln\le|\frac{R_0(z)R_0(w)+wz-1}{z-w}\ri| u(w)dw + \s(z)u(z)=|R_0(z)|,       \label{1stNDR}  \\
  \int_\G \ln\le|\frac{R_0(z)R_0(w)+wz-1}{z-w}\ri| v(w)dw + \s(z)v(z)=2z|R_0(z)| .  
  \label{2ndNDR}
  \end{align}
Note that the kernel in the NDR for the defNLS (dark) soliton gas represent ``half" of that for the   fNLS breather gas, see \eqref{dr_breather_gas1}-\eqref{dr_breather_gas2}.

\br \label{rem-err-est}
The error estimate in the transition from  \eqref{RBI-full} to its thermodynamic limit \eqref{1stNDR-full}   in the context of  fNLS breather gas was obtained in \cite{TW22}, subject to additional requirement that $\s(z)>0$ on $\G$. This result can be extended to defNLS soliton gases. 
It is generally assumed that this transition is justified  
for sufficiently smooth  $\s(z)\geq 0$ on $\G\subset [-1,1]$, in particular, for $\s\equiv 0$ that corresponds to soliton condensates. We will often use this assumptions in the rest of the paper without further comments.
\er

\br\label{rem-phase-shift}
We now want to check that the observation from Remark  \ref{rem-ph-shifts}  about the phase shifts of interacting dark solitons holds true for 
the 
1st NDR  \eqref{1stNDR} for defNLS soliton gases. Indeed, the above mentioned 
 phase shift is given (up to the sign) by equation (11) from \cite{CER21} (see also \cite{ZS}), as
\bea\label{Gena-shift}
\frac 1{2\sqrt {1-z^2}}\ln \le|\frac{(z-w)^2+(\sqrt{1-z^2}+\sqrt{1-w^2})^2}
{(z-w)^2+(\sqrt{1-z^2}-\sqrt{1-w^2})^2}\ri|=\cr
\frac 1{\sqrt {1-z^2}}\tanh^{-1}\frac{\sqrt{1-z^2}\sqrt{1-w^2}}
{1-zw}
\eea
It must coincide with the expression 
\be\label{phase-shift}
\frac 1{\sqrt {1-z^2}}\ln \le|\frac{-\sqrt{1-z^2}\sqrt{1-w^2}+wz-1}{z-w}\ri|
\ee
coming from the first NDR \eqref{1stNDR}.
Substitution $z={\cos} \z$, $w={\cos}  \x$ into both \eqref{phase-shift}, \eqref{Gena-shift} easily shows that they indeed  do coincide. 
\er

  \section{Solutions of the NDR}
  \label{sec-solution}
  In this section we discuss existence of solutions to the NDR \eqref{1stNDR}-\eqref{2ndNDR}  for defNLS soliton gas.
It turns out that these NDR can be reduced to that of the circular soliton gas for fNLS that was studied in the previous work of the authors \cite{TW2023}. A circular fNLS soliton gas has NDR  \eqref{NDR}-\eqref{NDR2}, where $\G^+$ belong to some semicircle in $\C^+$ centered at $z=0$. Let us write these NDR as
%
\begin{align}
	\int_{\tilde \G}\ln\le|\frac{Z-\bar W}{Z-W}\ri|U(W)|dW|+\tilde\sigma(Z)U(Z)&=\Im Z,\quad Z\in \tilde\G,\label{eq: NDR in circle}\\
	\int_{\tilde \G}\ln\le|\frac{Z-\bar W}{Z-W}\ri|V(W)|dW|+\tilde\sigma(Z)V(Z)&=-2\Im Z^2,\quad Z\in \tilde\G,\label{eq: NDR2 in circle}
\end{align}
where $\tilde\G\subset \{e^{i\alpha},\alpha\in (0,\pi)\}$. Connection between the solutions of the NDRs \eqref{1stNDR}-\eqref{2ndNDR} for defNLS soliton gas and  \eqref{eq: NDR in circle}-\eqref{eq: NDR2 in circle} for the circular fNLS gas is given through the Joukovski transformation 
\be\label{Jouk}
z=\hf\le( Z+\frac 1Z \ri)
\ee
as described in the following theorem.

\bt\label{thm-reduction} Functions
 $U(Z),V(Z)$ are 
  solutions of the  NDR  \eqref{eq: NDR in circle}-\eqref{eq: NDR2 in circle} for the fNLS circular gas if and only if
\begin{align}\label{eq: def-dos}
	u(z)=\frac{U(Z(z))}{\Im Z(z)},\quad
	v(z)=-\frac{V(Z(z))}{2\Im Z(z)}
\end{align}
are solutions of the NDR \eqref{1stNDR}-\eqref{2ndNDR}
for the 
defNLS (dark) soliton gas respectively,  where 
\begin{align}\label{eq: new sig}
	\sigma(z)=\tilde\sigma\le(Z\ri)\Im Z, \quad z\in (-1,1)
\end{align}
and $\G$ is the projection of $\tilde \G$ on $\R$.
\et
\begin{proof}
	In equation \eqref{1stNDR}, replace  $z=\Re Z$ and $w=\Re W$ with $Z,W\in \tilde\G\subset \{e^{i\alpha},\alpha\in (0,\pi)\}$. Let $\alpha,\theta$ be the corresponding arguments for $W,Z$. Then for the left hand side of \eqref{1stNDR} we have
	\begin{align*}
		&\int_\G \ln\le|\frac{R_0(z)R_0(w)+wz-1}{z-w}\ri| u(w)dw + \s(z)u(z)\\
		&=\int_{\tilde\G}\le|\frac{\cos\alpha\cos\theta-\sin\alpha\sin\theta-1}{\cos\theta-\cos\alpha}\ri|\le(u(\Re W)\Im W\ri)|dW|+\frac{\sigma(\Re Z)}{\Im Z}\le(u(\Re Z)\Im Z\ri)\\
		&=\int_{\tilde\G}\ln\le|\frac{-2\sin^2\frac{\theta+\alpha}{2}}{-2\sin\frac{\theta+\alpha}{2}\sin\frac{\theta-\alpha}{2}}\ri|U(W)|dW|+\tilde\sigma(Z)U(Z)
		\\&
		=\int_{\tilde\G}\ln\le|\frac{Z-\bar W}{Z-W}\ri|U(W)|dW|+\tilde\sigma(Z)U(Z),
	\end{align*}
where  we utilized equation \eqref{eq: new sig}. 
For the right hand side of \eqref{1stNDR}  we have
	\begin{align*}
		|R_0(z)|=\sin\theta=\Im Z.
	\end{align*}
	Thus we have shown that under the map $z=Re Z$, the first NDR for defNLS gas is transformed into the first NDR for the circular fNLS gas. For $z\in (-1,1)$ and $Z$
	in the upper unit semicircle, the map \eqref{Jouk} is bijective, whose inverse is given by
	\begin{align}\label{inv-Jouk}
		Z=z+i\sqrt{1-z^2},\quad z\in (-1,1),
	\end{align}
thus providing the one to one  correspondence between the solutions $u(z)$ and $U(Z)$ given by the first expression in \eqref{eq: def-dos}.

A direct computation shows
\begin{align*}
	-4z|R_0(z)|=-4\Re Z\Im Z=-2\Im Z^2.
\end{align*}
We can now repeat the same arguments for solutions $v(z)$ and $V(Z)$ for the second NDRs equations to complete the proof.

\end{proof}

 According to Theorem \ref{thm-reduction}, the existence, uniqueness  and properties  of the solutions to the NDR  \eqref{1stNDR}- \eqref{2ndNDR} follow from the corresponding properties of the NDR  \eqref{eq: NDR in circle}-\eqref{eq: NDR2 in circle} for the  fNLS circular gas. For fNLS soliton gases, questions of  existence and uniqueness were studied in \cite{KT2021}.
Based on these results
 one has:

 \bc
 Let $\sigma$ be continuous on $\G$ and the null-set $S_0=\{z\in\G|\sigma(z)=0\}$ be either empty or thick\footnote{S is thick (or non-thin) at $z_0$ if $z_0 \in\overline{ S \backslash \{z_0\}}$ and if, for every superharmonic function $u$ defined in a neighborhood of $z_0$, $\liminf_{S\backslash \{z_0\}\ni z\ra z_0}u(z)=u(z_0)$.} at each $z_0\in S_0$. Then the solution $u(z)$ to \eqref{1stNDR} exists and is unique; moreover, $u(z)\geq 0$ for $z\in \G$.  
 The same statements 
 are also valid for $v(z)$ except that $v(z)\in\R$ (it can be negative). 
 \ec
 
 \begin{proof}
 	Let $\tilde \G=Z(\G)$, where $\G\subset[-1,1]$ and the map $Z$ is given by \eqref{inv-Jouk}.
 Then a point $Z\in \tilde\G$ is a thick point for the null set of $\tilde \s$ if and only if the corresponding $z\in\G$ is a thick point for the null set of $\s(z)$.
 	According to 
 	Theorem 1.6 in \cite{KT2021}, the DOS $U(Z)$ for the circular fNLS exists,  is unique and  non-negative provided 
 	that the null set of $\tilde{\sigma}(Z)$ is thick at every point of $\tilde\G$.  In view of \eqref{eq: new sig}, the latter is true if and only if  the null set of  $\tilde \s$ is thick at every point of $\G$. Similar arguments work for the DOF $v(z)$.
 \end{proof}

 In the calculations below we reduce the NDR to its differentiated form, that can be expressed in terms of the the Finite Hilbert Transform (FHT) $H$. In certain cases, the latter expressions can be useful for explicit solutions $u,v$ of the NDR  \eqref{1stNDR}- \eqref{2ndNDR}.

 Direct calculations (see also \eqref{anti-der})  prove  the identity
 \be\label{diff-id}  
 \frac d{dz}  \ln\frac{R_0(z)R_0(w)+wz-\d^2}{{|z-w|}} =\frac {R_0(w)}{R_0(z)(w-z)},
 \ee  
 for any $\d\in\C$, where $R_0(z)=\sqrt{z^2-\d^2}$.  Then, the differentiated 1st NDR \eqref{1stNDR-full} and its temporal  counterpart become
 \begin{align}\label{1stNDR-full-diff}
 -\frac i{R_0(z)}\int_\G\frac{R_0(w)u(w)dw}{w-z}-i[\s(z)u(z)]'=\frac {-z}{R_0(z)},\\
 	\label{2ndNDR-full-diff}
 	-\frac i{R_0(z)}\int_\G\frac{R_0(w)v(w)dw}{w-z}-i[\s(z)v(z)]'=\frac {4z^2-2}{R_0(z)},
 \end{align}
 or
 \begin{align}\label{FTH-u}
 	\pi H[u(w)\sqrt{1-w^2}]+\sqrt{1-z^2} \frac d{dz}[\s(z)u(z)]=-z,\\
 	\pi H[v(w)\sqrt{1-w^2}]+\sqrt{1-z^2} \frac d{dz}[\s(z)v(z)]=4z^2-2.
 	\label{FTH-v}
 \end{align}
where $H$ denotes the FHT on $\G$.

\section{Solutions for the defNLS condensates}
\label{sec-condens}
 Next, we study the defNLS soliton condensate, i.e., the case of  $\sigma\equiv 0$, supported on the compact set  $\G\subset [-1,1]$. Then \eqref{FTH-u}- \eqref{FTH-v} become
 \begin{align}
	 	\pi H[u(w)|R_0(w)|](z)&=-z,\label{FHT-1}\\
	 	\pi H[v(w)|R_0(w)|](z)&=4z^2-2,\label{FHT-2}
	 \end{align}
 where $H$ denotes the finite Hilbert transform on $\G=\cup_{j=1}^nc_j$.  Here the  gaps $c_j=[a_{2j-1},a_{2j}],j=1,\cdots,n,$ where
  $-1=a_0\leq a_1\leq \dots\leq a_{2n}\leq a_{2n+1}=1$, 
 are formed by the accumulating  (sub-exponentially) shrinking  gaps in the thermodynamic limit.
 %
  The remaining intervals on  $[-1,1]$,  i.e., the intervals of $[-1,1]\setminus \cup_{j=1}^nc_j$, will be called bands $\g_j$, $j=1,\dots,n+1$. The band gap structure on $[-1,1]$ defines the hyperelliptic Riemann surface $\Rscr$, which we will refer to as ``superband" RS $\Rscr$. 
 \footnote{ If the branchcut of $R_0(z)$ is deformed from $[-1,1]$ to $[1,+\infty]\cup[-\infty,])$, the bands and gaps on $[-1,1]$  must be interchanged, i.e., bands become gaps and vice versa. The obtained band gap structure  would then coincide with that on the $p$-plane for circular fNLS gas from \cite{TW2023}.}

 Comparing with the circular fNLS soliton condensate, we can easily obtain the DOS $u(z)$ and DOF $v(z)$ for the defNLS soliton condensate, which we state as the following theorem. 
 \bt\label{thm condensate}
 The DOS $u(z)$ and DOF $v(z) $ for the defNLS soliton condensates are given by
 \begin{align}
	 	u(z)=\frac { P(z)}{R(z)R_0(z)},\qquad
	 	v(z)=\frac{  Q(z)}{R(z)R_0(z)},\label{eq: DOS+DOF}
	 \end{align}
 where {$u,v$  satisfy \eqref{FHT-1},  \eqref{FHT-2} respectively}, $R(z)=\le(\prod_{j=0}^{2n-1}(z-a_j)\ri)^{1/2}$ such that $R(z)\sim z^{n}$ as $z\ra \infty$ 
 and  $P,Q$  are polynomials of degree  $n+1$ and $n+2$  respectively such that $u,v$ satisfy the  band vanishing conditions:
  	\begin{align}\label{vanish}
	 	 		\int_{\g_j}u(z)dz=0,\quad \int_{\g_j}v(z)dz=0,\quad j=1,\cdots,n,
	 	\end{align}
	 	and $u(z)dz$, $v(z)dz$ are
	 	second kind meromorphic differentials on the Riemann surface $\mathcal{R}_n$.
 	 Moreover, $u,v\in \R$ on $\G$ and $u(z)>0$ for $z\in \G$.
\et
\begin{proof}
	The proof is the same as the one for the circular fNLS condensate, see \cite{TW2023}, which are based on solving the integral equations \eqref{FHT-1}, \eqref{FHT-2}, i.e., in inverting the FHT $H$.
Alternatively,
Theorem \ref{thm condensate} can be derived directly
 from Theorem 1.3 in \cite{TW2023} using 
 Theorem \ref{thm-reduction}.
\end{proof}

\br\label{rem-dp-infty}
It is not difficult to establish that the leading coefficient of $P$ is $\frac{1}{\pi}$. Thus,  Theorem \ref{thm condensate} states that 
\be\label{dp-infty}
dp_\infty=\pi u(z)dz
\ee
 is a real normalized quasimomentum differential on the Riemann surface $\Rscr$ defined by the $R(z)R_0(z)$.
Similar statement is true for  $v(z)$ and the quasienergy differential on $\Rscr$.
\er
In the rest of the section, we consider the defNLS condensates in the genus one case.  The underlying Riemann surface for the genus one condensate is shown as Fig.\ref{fig:g1}. The following Corollary follows from Theorem  \ref{thm condensate}.

\begin{figure}
	\centering
	\resizebox{0.8\textwidth}{!}{
		\tikzset{every picture/.style={line width=0.75pt}} 

\begin{tikzpicture}[x=0.75pt,y=0.75pt,yscale=-1,xscale=1]
	
	\draw  [dash pattern={on 4.5pt off 4.5pt}]  (186.83,134) -- (264.33,134) ;
	\draw [shift={(264.33,134)}, rotate = 0] [color={rgb, 255:red, 0; green, 0; blue, 0 }  ][fill={rgb, 255:red, 0; green, 0; blue, 0 }  ][line width=0.75]      (0, 0) circle [x radius= 1.34, y radius= 1.34]   ;
	\draw [shift={(186.83,134)}, rotate = 0] [color={rgb, 255:red, 0; green, 0; blue, 0 }  ][fill={rgb, 255:red, 0; green, 0; blue, 0 }  ][line width=0.75]      (0, 0) circle [x radius= 1.34, y radius= 1.34]   ;
	\draw    (264.33,134) -- (295,134) -- (341.83,134) ;
	\draw    (385.67,134) -- (424.33,134) ;
	\draw  [dash pattern={on 4.5pt off 4.5pt}]  (424.33,134) -- (469.83,134) ;
	\draw [shift={(469.83,134)}, rotate = 0] [color={rgb, 255:red, 0; green, 0; blue, 0 }  ][fill={rgb, 255:red, 0; green, 0; blue, 0 }  ][line width=0.75]      (0, 0) circle [x radius= 1.34, y radius= 1.34]   ;
	\draw [shift={(424.33,134)}, rotate = 0] [color={rgb, 255:red, 0; green, 0; blue, 0 }  ][fill={rgb, 255:red, 0; green, 0; blue, 0 }  ][line width=0.75]      (0, 0) circle [x radius= 1.34, y radius= 1.34]   ;
	\draw  [dash pattern={on 4.5pt off 4.5pt}]  (341.83,134) -- (385.67,134) ;
	\draw [shift={(385.67,134)}, rotate = 0] [color={rgb, 255:red, 0; green, 0; blue, 0 }  ][fill={rgb, 255:red, 0; green, 0; blue, 0 }  ][line width=0.75]      (0, 0) circle [x radius= 1.34, y radius= 1.34]   ;
	\draw [shift={(341.83,134)}, rotate = 0] [color={rgb, 255:red, 0; green, 0; blue, 0 }  ][fill={rgb, 255:red, 0; green, 0; blue, 0 }  ][line width=0.75]      (0, 0) circle [x radius= 1.34, y radius= 1.34]   ;
	
	\draw (257.33,135.4) node [anchor=north west][inner sep=0.75pt]  [font=\small]  {$a_{1}$};
	\draw (337.33,135.4) node [anchor=north west][inner sep=0.75pt]  [font=\small]  {$a_{2}$};
	\draw (378,135.4) node [anchor=north west][inner sep=0.75pt]  [font=\small]  {$a_{3}$};
	\draw (417,135.4) node [anchor=north west][inner sep=0.75pt]  [font=\small]  {$a_{4}$};
	\draw (168.07,135.28) node [anchor=north west][inner sep=0.75pt]  [font=\small,rotate=0]  {$a_{0} =-1$};
	\draw (455.53,135.16) node [anchor=north west][inner sep=0.75pt]  [font=\small,rotate=0]  {$a_{5} =+1$};

\end{tikzpicture}	}
	
	\caption{Riemann surface for the genus one defNLS gas. The dashed lines indicate the accumulation of the gaps: $(-1,a_1)$, $(a_2,a_3)$ and $(a_4,1)$. }
	\label{fig:g1}
\end{figure}


\bc
\label{cor:genus one}
Given the contour $\G$ as shown in Fig.\ref{fig:g1}, the solution $u(z), v(z) $ to the differentiated NDR \eqref{FHT-1},  \eqref{FHT-2}  for genus one defNLS condensate satisfying conditions \eqref{vanish}
is given by
\begin{align}
	u(z)&=u_1(z;a_1,a_2,a_3,a_4)=\frac{1}{\pi}\left(\frac{z^2-\frac{1}{2}l_1z+A}{R(z)}\right),\label{g1-DOS}\\
	v(z)&=v_1(z;a_1,a_2,a_3,a_4)=\frac{1}{\pi}\left(\frac{4z^3-2l_1z^2+(2l_2-l_1^2/2)z+B}{R(z)}\right), \label{g1-DOS-v}
\end{align}
where
\begin{align}
	R(z)&=\sqrt{\prod_{j=1}^4(z-a_j)}, ~~~	l_1=\sum_{j=1}^4a_j, ~~~l_2=\sum_{1\leq i<j\leq 4 }a_ia_j,\\
	A&=-\frac{E(m)}{2K(m)}(a_3-a_1)(a_4-a_2)+\frac{1}{2}(a_3a_4+a_2a_1),\\
	B&=-\frac{E(m)}{2K(m)}(a_3-a_1)(a_4-a_2)l_1+\frac{1}{2}(a_4a_3-a_2a_1)(a_4+a_3-a_2-a_1),\\
	m&=\frac{(a_4-a_3)(a_2-a_1)}{(a_4-a_2)(a_3-a_1)},
\end{align}
and, the branch of $R(z)$ is chosen so that $R(z)\sim z^2$ as ${z\ra \infty}$.
Here $E(m),K(m)$ are complete elliptic integrals  of the first and the second kind respectively.
The effective velocity is then given by
\begin{align}
	s(z)=s_1(z;a_1,a_2,a_3,a_4)=4z-\frac{Cz+B}{z^2-\frac{1}{2}l_1z+A},
\label{g1-effv}
\end{align}
where
\begin{align}
	C=-\frac{2E(m)}{K(m)}(a_3-a_1)(a_4-a_2)+\frac{1}{2}(a_3+a_4-a_2-a_1)^2.
\end{align}
\ec

\begin{proof}
	The results follow from the Corollary 3.1 in \cite{TW22} and Theorem \ref{thm-reduction}. They can also be verified directly.
\end{proof}

\begin{figure}[ht]
	\centering
	\begin{minipage}{0.45\textwidth}
		\centering
		\includegraphics[width=\linewidth]{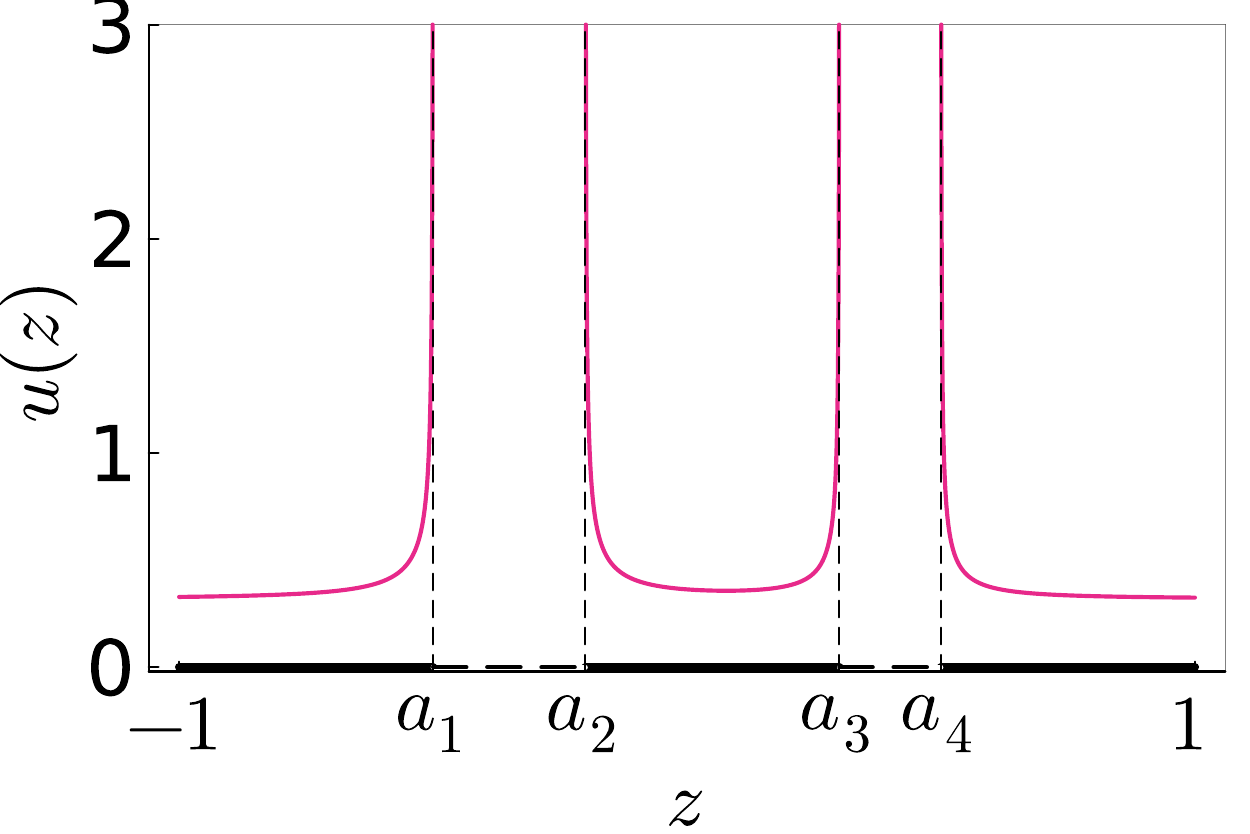}
	\end{minipage}\hfill
	\begin{minipage}{0.45\textwidth}
		\centering
		\includegraphics[width=\linewidth]{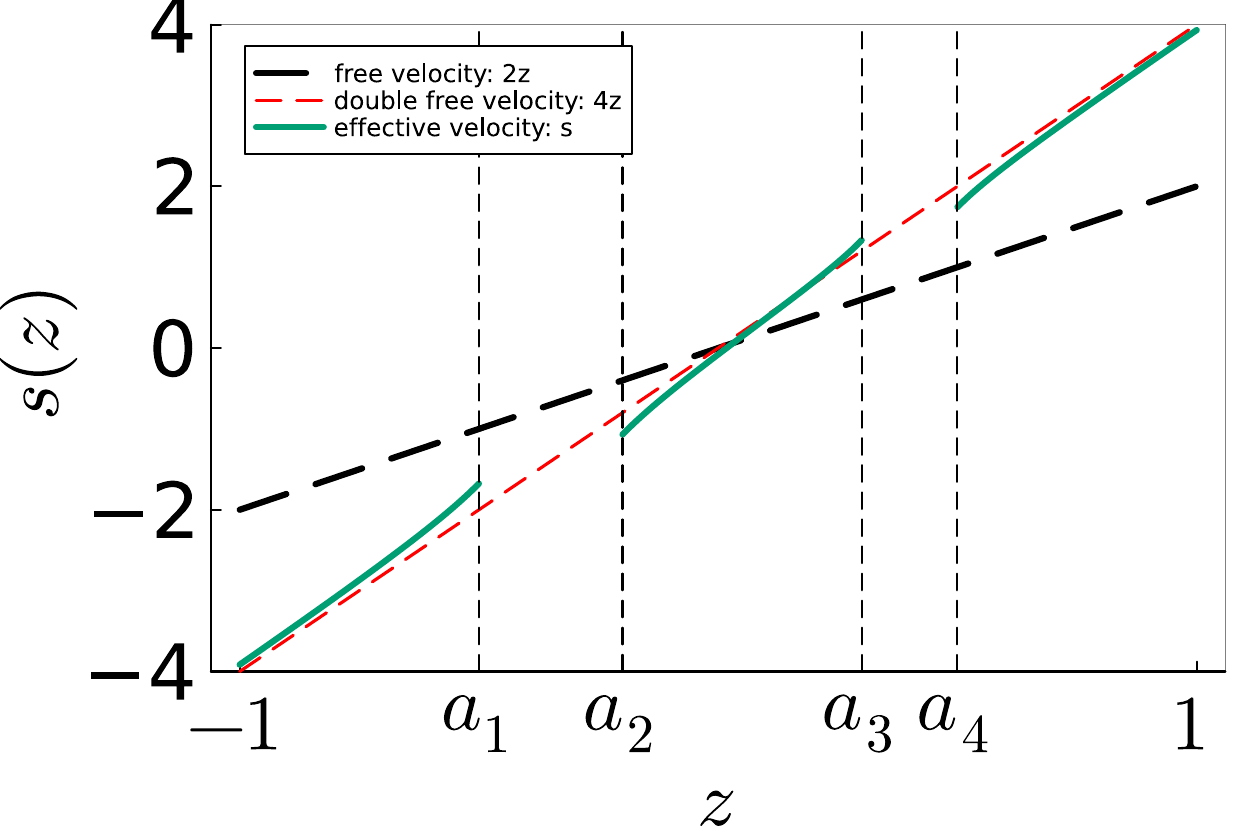}
	\end{minipage}
	\caption{\textbf{Left}: A plot of DOS $u(z)$ whose support is on $[-1,a_1]\cup [a_2,a_3]\cup [a_4,1]$. 
		 \textbf{Right}: A plot of effective velocity $s(z)$ which is supported on $[-1,a_1]\cup [a_2,a_3]\cup [a_4,1]$. As shown in the plot, the effective velocity $s(z)$ is approximately equal to twice the free velocity when the gaps become small. Similar velocity doubling phenomenon for fNLS circular condensate was established  in \cite{ET2020}, \cite{TW2023} and experimentally  observed in \cite{Circular}.}
\end{figure}

\bc
\label{g0-master case}

When
$a_4=a_3$,  the solution \eqref{g1-DOS} and \eqref{g1-DOS-v} reduce to
\begin{align}
	u(z)&=u_0(z;a_1,a_2)=\frac{1}{\pi}\frac{(z-\frac{1}{2}(a_2+a_1))}{\sqrt{(z-a_2)(z-a_1)}},\label{g0-DOS}\\
	v(z)&=v_0(z;a_1,a_2)=\frac{4}{\pi}\frac{(z^2-\frac{1}{2}(a_2+a_1)z-\frac{1}{8}(a_1-a_2)^2)}{\sqrt{(z-a_2)(z-a_1)}},\label{g0-DOS-v}
\end{align} 
where $z\in [-1,a_1]\cup [a_2,1]$, and the radical $\sqrt{(z-a_1)(z-a_2)}$ is non-negative on $[a_2,1]$ and non-positive on $[-1,a_1]$. The effective velocity is then given by
\begin{align}
	s(z)=s_0(z;a_1,a_2)=4z-\frac{\frac{1}{2}(a_2-a_1)^2}{z-(a_2+a_1)/2}.\label{g0-effv}
\end{align}

\ec
\begin{proof}
	Using solutions \eqref{g1-DOS} and \eqref{g1-DOS-v}, taking the limit as $a_3\ra a_4-$, one obtains \eqref{g0-DOS} and \eqref{g0-DOS-v} respectively.
\end{proof}

\bc\label{cor-vac}
If in the conditions of Corollary \ref{g0-master case} we have $a_1=a_2$, then
\be\label{vac-param}
u_0(z)=\frac 1\pi, \quad v_0(z)=\frac {4z}\pi\quad \text{and} \quad s_0(z)=4z.
\ee
\ec

We note that the effective velocity in the defNLS condensate with the spectral support $\G=[-1,1]$
is twice the velocity of a free soliton. The same holds true for fNLS circular soliton condensates when the spectral support set $\G^+$ in $\C^+$ is the upper semicircle (\cite{ET2020}).

\section{Modulational dynamics for the defNLS soliton gases}
\label{sec-kinetic}
If we have a nonhomogeneous  defNLS soliton gas the bands evolve with  $x,t$ according to 
\begin{align}
	u_t+v_x=0,\label{eq: cons-law}
\end{align}
where $u,v$ are given by \eqref{eq: DOS+DOF}. The following theorem converts \eqref{eq: cons-law} into a system of  PDEs  for the end points $a_j$ of the bands of $\G$ (branch points for the corresponding Riemann surface $\Rscr$), also known as modulation equations.
The proof of this theorem is based on the same arguments as those used in \cite{CERT}, \cite{TW2023} for  KdV and fNLS  nonhomogeneous  soliton condensates respectively. For convenience of the reader, we present this proof
below.

\bt\label{th-kin}
For a nonhomogeneous  defNLS soliton condensate with distinct endpoints $a_j$,  the equation
\eqref{eq: cons-law} is equivalent to  the system of modulation equations
\begin{align}
	\partial_t a_j(x,t)+V_j\partial_x a_j(x,t)=0,\quad  j=1,\cdots,2n\label{eq: mod eq}
\end{align}
where
\begin{align}
	\left. V_j=\frac{Q(z)}{P(z)}\ri\vert_{z=a_j}, \qquad j=1,\dots,2n,\label{eq: V velocity}
\end{align}
and the polynomials $P,Q$ are defined in Theorem \ref{thm condensate}.
\et
\begin{proof}
	Let us first show that equation \eqref{eq: cons-law} implies \eqref{eq: mod eq}. Using 
	\eqref{eq: DOS+DOF}, we can write
	\eqref{eq: cons-law}  as
	\begin{align}
		\le(\frac{P}{RR_0}\ri)_t+\le(\frac{Q}{RR_0}\ri)_x=0\qquad\text{or}\qquad 	\le(\frac{P}{R}\ri)_t+\le(\frac{Q}{R}\ri)_x=0,
	\end{align}
	which can be further written as
	\begin{align}
		P_t+Q_x=P(\log(R))_t+Q(\log(R))_x.\label{eq:PQ}
	\end{align}
	Multiplying by $(z-a_j)$  both sides of \eqref{eq:PQ} and taking  limit $z\ra a_j$, one obtains
	\begin{align}
		P(a_j)(a_j)_t+Q(a_j)(a_j)_x=0.\label{eq:PQV}
	\end{align}
	Note the band vanishing conditions \eqref{vanish}
	imply that  zeros of $P$ are located on the bands (one in each). Thus $P(a_j)\neq 0$.  Dividing  \eqref{eq:PQV}  by $P(a_j)$ one gets equation \eqref{eq: mod eq},  where $V_j$ given by \eqref{eq: V velocity} is bounded.
	
	Assume now the modulation equation for the endpoints $\{a_j,j=1,\cdots,2n\}$ are given by \eqref{eq: mod eq}. Following the argument in \cite{FFM} (or the authors' previous work \cite{TW2023}), we consider the  differential
	\begin{align}
		\Omega=u_tdz+v_xdz
	\end{align}
on the superband RS $\Rscr$,
	where $u,v$ are given by Theorem \ref{thm condensate}. Then 
	\begin{align}
		\Omega=\frac{1}{2}\le[\sum_{j=1}^{2N}\frac{P(z)(a_j)_t+Q(z)(a_j)_x}{z-a_j}\frac{dz}{R(z)R_0(z)}\ri]+\frac{P_t(z)+Q_x(z)}{R(z)R_0(z)}dz,
	\end{align}
	which implies that $\Omega$ has at most double poles at $a_j$. But the equation \eqref{eq: mod eq} implies $\Omega$ is holomorphic at $a_j$. Note again by Theorem \ref{thm condensate}, we observe that $\Omega$ is holomorphic at $z=\infty$, thus, $\Omega$ is a holomorphic differential on $\Rscr$. Since by \eqref{vanish} all the band integrals
	  of $udz$ and $vdz$ are  zero, so are their $x,t$ derivatives. We thus get that $\Omega=0$, which implies \eqref{eq: cons-law}.

\end{proof}

One can interpret Theorem \ref{th-kin} as saying that the kinetic equation for defNLS soliton condensate is equivalent to  defNLS-Whitham equations  (\cite{Kodama}) for the dynamics of  endpoints of spectral bands  of slowly modulated finite gap defNLS solutions.

In the rest of the section, we study the time evolution of a nonhomogeneous DOS $u$ of the defNLS soliton condensate of genus zero, which is defined by the initial distribution of $a_1(x,t)$ at $t=0$. Without loss of generality, we set $a_2=1$ and the branch point $a_1$ is the only moving point. According to Theorem \ref{th-kin}, the dynamics of $a_1$ is governed by
\begin{align}
	\partial_t a_1+V_1 \partial_xa_1=0,\quad V_1=3a_1+1,
	\label{burges eq}
\end{align}
where the latter expression is obtained by calculating
$s_0(a_1;a_1,1)$ from \eqref{g0-effv}. This equation coincides (up to a factor due to the different set-ups for the defocusing NLS equation) with the Whitham equation (2.9) in \cite{Kodama}. It also coincides (up to a factor) with the Whitham equation  (1.16a) in \cite{JinLev} for $\hat\lambda_-$, where $\hat\lambda_+=1$ is fixed. 

The classical Riemann problem consists of finding solution to a system of hyperbolic conservation laws, like \eqref{eq: mod eq}, subject to piecewise-constant initial conditions exhibiting discontinuity, say, at $x=0$. The
solution of such Riemann problem generally represents a combination of constant states, simple (rarefaction) waves and strong discontinuities (shocks or contact discontinuities). 

\br
The modulation equations 
	\eqref{eq: mod eq} form a strictly hyperbolic system of first order quasi-linear PDEs provided that all the branch points are distinct (otherwise it will be just hyperbolic).  This system is in the diagonal  (Riemann) form with all the coefficients (velocities) being real.  The Cauchy data for this system consists of   $\vec a(x,0)$, which denotes the vector of all the endpoints $a_j(x,0)$.  The system has a unique local (classic) real solution provided $\vec a(x,0)$ is of $C^1$ class and real and the coefficients $V_j(\vec a)$ are smooth and real (see for example see \cite{DafBook}, Theorem 7.8.1). 
	\er
	
\br\label{rem-break}

As it is well known, hyperbolic system may develop singularities in the $x,t$ plane, which, in the case of modulation equations  \eqref{eq: mod eq}, lead to collapse of a band or a gap, or to appearance of a new  ``double point" that will open into a band or gap. In any case,  at a point of singularity (also known as a breaking point), two or more endpoints from $\vec a(x,t)$ collide or a new pair of collapsed  double points appear, so that the Riemann surface $\mathcal R$ develop a singularity. This type of situation was observed and discussed in a variety of different settings,
see, for example, \cite{TVZ04}, \cite{Bertola}, including in slowly modulated  finite gap defNLS solutions \cite{Kodama}.
Technically, Theorem \ref{th-kin} is not applicable at a breaking point, but, according to general results of \cite{BT14} (and also \cite{Kodama}), solutions of the modulation equations \eqref{eq: mod eq} can be continued into regions of different genera  beyond  breaking  points.
 Some examples of the evolution of the defNLS  condensate after the breaking point will be discussed in the rest of the section.

\er

Let us consider the Riemann problem for a genus 0 defNLS condensate determined  by the step initial data
\begin{align}
	a_1(x,t=0)=\begin{cases}
		q_-,& x<0\\
		q_+,& x>0
	\end{cases},\quad  q_+\neq q_-,\label{eq:a1 ini}
\end{align}
where $q_\pm\in (-1,1)$.

There are two cases depending on whether $q_->q_+$ or $q_-<q_+$. In the first case, it is well known (see \cite{Kodama}) that the equation \eqref{burges eq} admits a unique global solution (a rarefaction wave).  In the second case, higher genus Riemann surfaces are need to describe the evolution of the nonhomogeneous defNLS gas. In this section, we will restrict ourselves to the genus 0 and genus 1 cases. And following similar arguments in \cite{CERT,Kodama,TW2023}, we give explicit formulae of DOS/DOF for the nonhomogeneous defNLS gas.

\subsubsection*{Rarefaction wave: $q_-<q_+$}
In this case, the initial data for $a_1$ is given by \eqref{eq:a1 ini} with $q_->q_+$. Then the equation \eqref{burges eq} admits the following rarefaction wave solution:

 For $(x,t)\in \R\times \R_+$,
	\begin{align}
		a_1(x,t)=\begin{cases}
			q_-,& x\leq V_{1-}t,\\
			-\frac{x}{3t}-\frac{1}{3},& V_{1-}t<x<V_{1+}t,\\
			q_+,& x\geq V_{1+}t,
		\end{cases}
	\end{align}
	where $V_{1\pm}:=\lim_{a_1\ra q_\pm}=3q_\pm+1$.
	
	Accordingly, the DOS and DOF for the defNLS condensates are given by
	\begin{align}
		u(z;x,t)=u_0(z;a_1(x,t),1)=\frac{1}{\pi}\frac{z-\frac{1+a_1(x,t)}{2}}{\sqrt{(z-a_1(x,t))(z-1)}},
	\end{align}
	and 
	\begin{align}
		v(z;x,t)=v_0(z;a_1(x,t),1)=\frac{4}{\pi}\frac{\le(z^2-\frac{1}{2}(1+a_1(x,t))z-\frac{1}{8}(a_1(x,t)-1)^2\ri)}{\sqrt{(z-a_1(x,t))(z-1)}}.
	\end{align}
	
	A density plot of the DOS is shown in Fig.\ref{fig:density plot DOS}(a), and the dashed line in the plot shows the graph of $a_1(x,t)$.
	
\subsubsection*{Dispersive shock wave: $q_->q_+$}
 In this case, a wave-breaking occurs immediately, thus we need higher genus DOS to regularize the singularities (see for example in \cite{CERT} for the KdV condensates). Such regularization is well-known in literature of KdV dispersive shock wave problem. In fact, one can find the following DOS and DOF that regularize the singularities:
 \begin{align}
 	u(z;x,t)&=\begin{cases}
 		u_0(z;q_-,1),& x<V_{2+}t,\\
 		u_1(z;q_+,a_2(x,t),q_-,1), & V_{2-}t<x<V_{2+}t,\\
 		u_0(z;q_+,1),& V_{2-}t<x,
 	\end{cases}\\
 	v(z;x,t)&=\begin{cases}
 		v_0(z;q_-,1),& x<V_{2+}t,\\
 		v_1(z;q_+,a_2(x,t),q_-,1), & V_{2-}t<x<V_{2+}t,\\
 		v_0(z;q_+,1),& V_{2-}t<x,
 	\end{cases}
 \end{align}
 where $a_2$ is the only moving branching point, which is governed (similar to \eqref{burges eq}) by
 \begin{align}
 	\partial_t a_2+V_2\partial_x a_2=0,\quad V_2=s_1(z;q_+,a_2,q_-,1)|_{z=a_2},
 \end{align}
 where $s_1$ is defined by \eqref{g1-effv}. Consider the self-similar reduction, the dynamic of $a_2$ is implicitly determined by the following equation:
 \begin{align}
 	\frac{x}{t}=V_2:=(q_++q_-+1+a_2)-\frac{2(a_2-q_+)(a_2-q_-)}{(q_+-q_-)\mu(m)-(a_2-q_-)},\label{eq: g1 vel}
 \end{align}
 where 
 \begin{align}
 	m&=\frac{(1-q_-)(a_2-q_+)}{(q_--q_+)(1-a_2)},\quad 
 	\mu(m)=\frac{E(m)}{K(m)}, \label{m,mu}.
 \end{align}
 And the critical velocities $V_{2\pm}:=\lim_{a_2\ra q_\pm}V_{2}$ are explicitly given by
 \begin{align}
 	V_{2+}&=\le(\frac{8q_+^2-4q_+q_--q_-^2-4q_++2q_--1}{2q_+-q_--1}\ri),\label{Def: V2pm}\\
 	V_{2-}&=1+2q_-+q_+.
 \end{align}
 
\begin{figure}[!ht]
	\subfloat[]{%
		\includegraphics[width=0.48\textwidth]{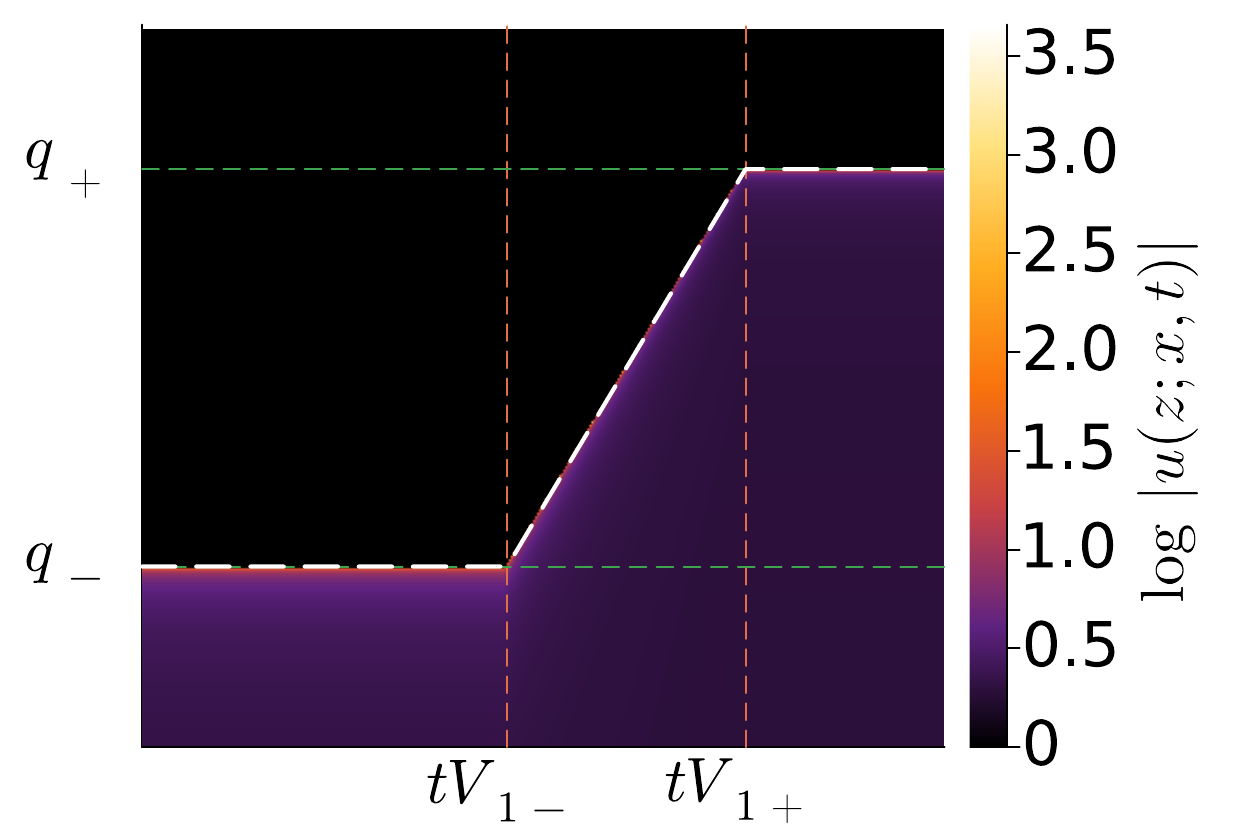}
	}
	\hfill
	\subfloat[]{%
		\includegraphics[width=0.48\textwidth]{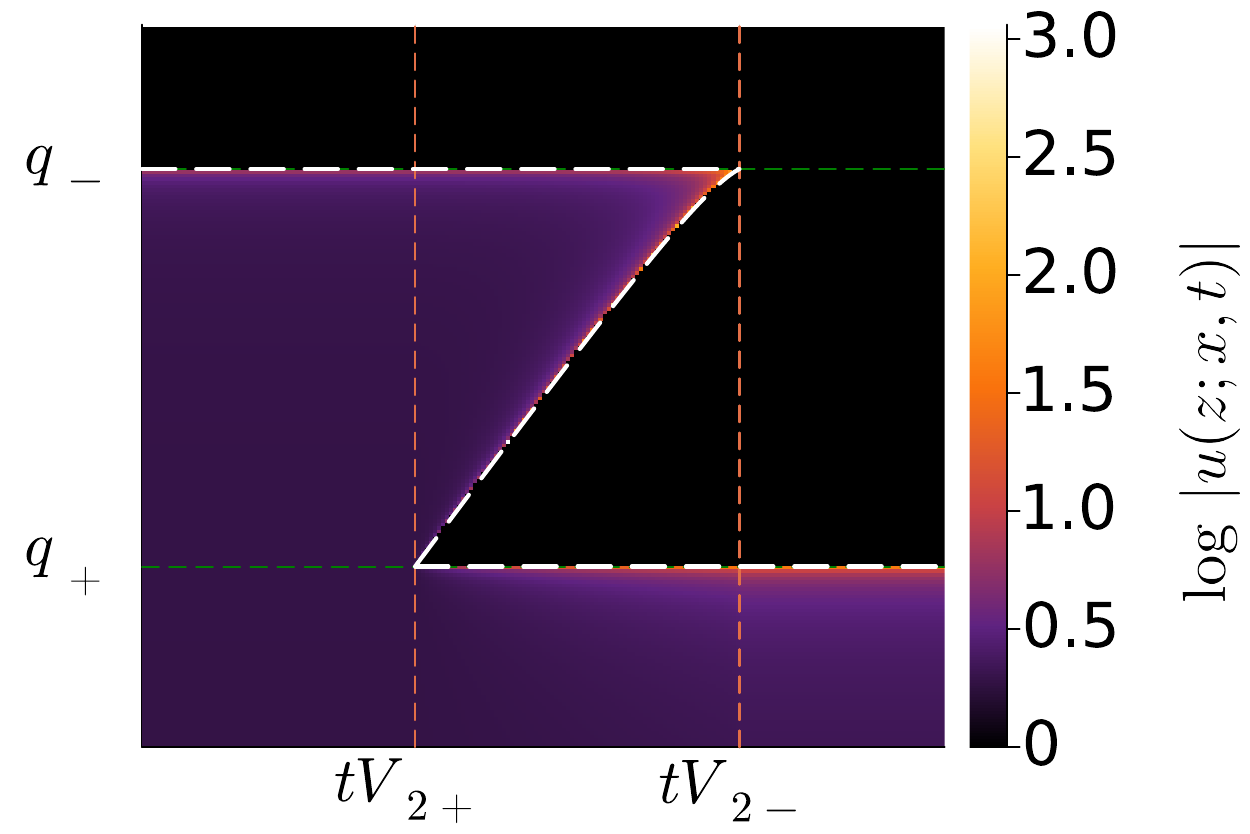}
	}
	\caption{Solutions to the kinetic equation for the defNLS  condensate with step function initial data (Riemann problem) \eqref{eq:a1 ini}. (a) Rarefaction wave solution when $q_-<q_+$; Dashed line: $a_1(x,t)$;  Colors: the values of DOS $u_0(z;a_1(x,t),1)$;(b) Dispersive shock wave solution when $q_->q_+$. Dashed line: $a_2(x,t)$; Colors: {the log value of DOS} $u_1(z;q_+,a_2(x,t),q_-,1)$.}
	\label{fig:density plot DOS}
\end{figure}

 It is easy to check that $V_{2+}<V_{2-}$, which explains why a wave-break occurs immediately. A density plot of the genus one DOS is plot in Fig.\ref{fig:density plot DOS}(b),  and the dashed line in the plot shows the graph of $a_2(x,t)$.

\section{Thermodynamic limit for quasimomentum differentials}
\label{sec-quasi}

The thermodynamic limit of the quasimomentum differential 
\be\label{dpNf}
\frac{dp_N}{dz}= 1  -\sum_{j=1}^\infty \frac{I_{m,N} }{z^{m+1}}     
\ee
 for finite gap solutions for the fNLS defined on $\Rscr_N$ was  introduced  in \cite{TW22}  as
\be\label{dp}
\frac{dp}{dz}:= 1  -\sum_{m=1}^\infty \frac{I_{m }}{z^{m+1}},     
\ee
where  $I_m=\lim_{N\ra\infty} I_{m,N}$, $m\in\N$.  In the assumption that the length of each band except possibly one does not exceed some $\d>0$, the coefficients $I_{m,N}$ were 
 shown (Theorem 1.1, \cite{TW22}) to have the form
\begin{align}\label{br-avg-den}
	I_{m,N}=\le[\frac{m}{2\pi i}\sum_{|j|=1}^N U_{j,N}\oint_{{\bf B}_j}\frac{[\z^{m-1}R_0(\z)]_+d\z}{R_0(\z)}+md_m\d_0^{m+1}\ri](1+O(N^\hf\delta)),
\end{align}
where: 
$R_0(z)=(z^2-\d_0^2)^\hf$ that behaves like $z+o(1)$ as $z\ra\infty$ with $\d_0\in i\R^+$; the polynomial $[f(\z)]_+$ consists of the non-negative powers   of the Laurent series expansion of $f(\z)$ at infinity;  $U_{j,N}=\hf k_j=\hf\oint_{{\bf A}_j}dp_N$ with ${\bf A}$-cycles being positively/negatively oriented small loops around each shrinking band in $\C^\pm$ respectively,  and; each
${\bf B}_j$ is the ${\bf B}$ cycle on $\Rscr_{N}$ intersecting $[-\d_0,\d_0]$ and ${\bf A}_j$ and oriented accordingly.
%
Additionally, 
$d_m\d_0^{m+1}$, $m\in\N$,
denote the coefficient of $z^{-m}$ in the Laurent expansion of $R_0(z)$ at infinity, that is,
$d_m=0$ when $m$ is even and 
\be\label{dm-int}
d_m= -\frac{1}{m}\frac{m!!}{(m+1)!!}
\ee 
when $m$ is odd.  

Based on the expansion \eqref{dp}, different  expressions for  $\frac{dp}{dz}$ in terms of the DOS $u(z)$ were derived for the cases of fNLS soliton and breather gases. In particular, 
in Theorem 3.18 of \cite{TW22} it was proven that 
\begin{align}\label{2gx-brea}
	2g_x(z)=i\int_\G u(\z)\ln\frac{R_0(\z)R_0(z)+\z z-\d_0^2}{\z-z} |d\z|+{z-R_0(z)},   
\end{align}
where $z\in \bar\C\setminus\G$
and $ \frac{dp}{dz}=1-2g_{xz}(z)$. 

In view of  \eqref{diff-id}, one has
\be\label{2gxz-brea}
2g_{xz}(z)=\frac i{R_0(z)}\int_\G 
\frac{u(\z)R_0(\z)}{\z-z} |d\z|+1-\frac z{R_0(z)}, 
\ee
so that
\be\label{quasi-bre}
\frac{dp}{dz}=\frac z{R_0(z)}-\frac i{R_0(z)}\int_\G 
\frac{u(\z)R_0(\z)}{\z-z} |d\z|=\frac z{R_0(z)}+\frac {2\pi}{R_0(z)}C_\G[\breve u R_0](z),
\ee
which is the thermodynamic limit of the quasimomentum differential for the breather gas with DOS $u(z)$ on $\G$, where $u(\z)|d\z|=\breve u(\z)d\z$. 

Let us make sure   that  \eqref{quasi-bre}  in the limit $\d\ra 0$ is consistent 
with the expression for $\frac{dp}{dz}$ for the fNLS soliton gas obtained in Theorem 3.15  \cite{TW22}. Indeed, using subscripts ``bre" and ``sol" to distinguish the corresponding quasimomentum differentials and observing  $R_0(z)\ra z$ as $\d\ra 0$, in this limit one has
\be\label{lim-del-0}
\le(\frac{dp}{dz}\ri)_{bre}=
\frac z{R_0(z)}+\frac {2\pi}{R_0(z)}C_\G[\breve u R_0](z) \ra 1+{2\pi}C_\G[\breve u ](z)=	\le(\frac{dp}{dz}\ri)_{sol},
\ee
which follows from
\be
\frac 1z \int_\G 
\frac{u(\z)\z}{\z-z} |d\z|=\frac 1z \int_\G u(\z)|d\z|+\int_\G 
\frac{u(\z)}{\z-z} |d\z|=\int_\G 
\frac{u(\z)}{\z-z} |d\z|
\ee
since $u(z)$ is real and anti Schwarz symmetrical.

\subsection{Defocusing NLS}
Let us go back to defNLS soliton gas. Our current goal is to show that the formula \eqref{quasi-bre} for  the thermodynamic limit of the quasimomentum differential for fNLS breather gas {can be adjusted} for the   defNLS (dark) soliton gas, where we take $R_0(z)=(z^2-1)^\hf$ and $\G$ to be the accumulation set of the  centers of shrinking bands
in the thermodynamic limit.

 The first step is to verify that the expression \eqref{br-avg-den} is valid for the quasimomentum differential $dp_N$ for the Riemann surface $\Rscr_N$ for defNLS finite gap solutions, which was  introduced in \eqref{dpN}.
 Expression \eqref{br-avg-den} for fNLS
 was derived by considering the $g$-function for finite gap solutions on $\Rscr_N$. This procedure is similar for defNLS (\cite{Zhou}),
where the
 Riemann surface $\Rscr_N$  has  $N+1$ branchcuts $\g_{j,N}
 \subset [-1,1]$. The g-function $g_x$ is defined by the RHP
	\be
	g_{x+}(z)+g_{x-}(z)=z+W_j, \quad z\in\g_{j,N}, \label{gfct for u}
	\ee
where $g_x(z)$ must be analytic in $\bar\C\setminus \cup_{j=1}^{N+1}\g_{j,N}$, $W_{N+1}=0$
and the real constants $W_j$, $j=1,\dots,N$ are to be determined by the requirement that $g_x$ is analytic at $z=\infty$. Then, by Sokhotski-Plemelj formula,

\bea\label{gform3}
	2g_x(z)=\frac{\tilde R(z)}{2\pi i}
	\sum_{j=1}^{N+1} \oint_{\g_{j,N}}{\frac{[\z+W_j]d\z}{(\z-z)\tilde R(\z)}}=\cr
	{\frac{\tilde R(z)}{2\pi i}}\le[	\sum_{j=1}^{N+1} \oint_{\g_{j,N}}{\frac{\z d\z}{(\z-z)\tilde R(\z)}}+
	\sum_{j=1}^{N+1}U_{j,N} \oint_{{\bf B}_j}\frac{d\z}{(\z-z)\tilde R(\z)}\ri],\label{gx defnls}
\eea
where $\tilde R(z)=R_0(z)R(z)$, see \eqref{R},
$z$ is outside of each loop and, as in \cite{ET2020},  $U_{j,N}:=W_{j+1}-W_j=\hf k_j$ (see \eqref{waven-freq} and the definition of ${\bf A}$ cycles below equation \eqref{appr-omega_A}).
Equation \eqref{gform3} has the same form as the expression for the 
g-function for the fNLS in \cite{TW22}, and the thermodynamic limit of   integrals over $\bf B$ cycles and the scaling in both situations are the same.
Therefore,  expressions \eqref{br-avg-den} for the defNLS soliton gases reads:
\begin{align}
	I_{m,N}=\le[\frac{m}{2\pi i}\sum_{j=1}^N U_{j,N}\oint_{{\bf B}_j}\frac{[\z^{m-1}R_0(\z)]_+d\z}{R_0(\z)}+md_m\d_0^{m+1}\ri](1+O(N^\hf\delta)).\label{defNLS dpN}	
\end{align}

Repeating the arguments of Theorem 3.18 from \cite{TW22} with the limiting spectral support  $\G=\cup_{j=1}^n c_{j,n}$
to show 
that  \eqref{quasi-bre}  coincides with  $\frac{dp}{dz}$ for the defNLS soliton gas,  defined by \eqref{dp}.
  
\bp\label{theo-ImN} 
The thermodynamic limit $\frac{dp}{dz}$ of quasimomentum differentials $\frac{dp_N}{dz}$ from \eqref{dpN} is given by \eqref{quasi-bre} with $\G=\cup_{j=1}^n c_{j,n}$ and $R_0=(z^2-1)^\hf\sim z$ as $z\ra\infty$.
   \ep

Let us apply Proposition \ref{theo-ImN}  to the genus 0 condensate with  $\G=[-1,1]$.    According to \eqref{vac-param}
$u(z)=\frac 1\pi$. Substituting that into \eqref{quasi-bre} with $R_0(z)=\sqrt{z^2-1}$ and using
\be\label{Hil-sq}
\int_{-1}^1 
\frac{u(\z)R_0(\z)d\z}{\z-z} =\frac 1{2\pi}\oint \frac{R_0(\z)d\z}{\z-z} =i(R_0(z)-z),
\ee
 we obtain 
\be\label{vac}
\frac{	dp}{dz}=1.
\ee
That means that $\langle |\psi|^2\rangle$, as well as the higher average conserved densities for  realizations of such condensate  are zeroes,  see Section \ref{sec-kurt} below,
i.e., the genus zero defNLS soliton condensate  with spectral support $\G$  coinciding with the background is vacuous.

Let us summarize some results from  \cite{TW22} about the thermodynamic limit  $\frac{dp}{dz}= \lim_{N\ra\infty} \frac{dp_N}{dz}$, see Section \ref{sec-intro},
of the quasimomentum differential $dp_N$ for fNLS soliton and
 breather gases.

		\br It is interesting to calculate  $\frac{dp}{dz}, \frac{dq}{dz}$ for the fNLS Akhmediev breather condensate with $\g_0=[-i,i]$ and $\G$ coinciding with the left shore of $\g_0$ oriented up. It  easily follows  from \eqref{dr_breather_gas1} that $u(z)\equiv 0$ (in fact, true for any $\G^+\subset [0,i]$). The second NDR \eqref{dr_breather_gas2} after some calculations yields $v(z)=\frac{4z}{i\pi}$. Substituting $u,v$ into \eqref{quasi-bre}, \eqref{eq: quasi-eng}, one obtains
	\be\label{pq-Akh} 
	\frac{dp}{dz}=\frac{z}{R_0(z)}=1+\frac 1{2z^2}+\frac 3{8z^4}+\dots,\qquad
	\frac{dq}{dz}=2R_0(z)-\frac{2z^2+2}{R_0(z)}+4z=4z.
	\ee
	In light of the  Section \ref{sec-kurt} below, we calculate that  for this fNLS breather gas the kurtosis $\kappa=2$. 
\er

	\bp\label{prop:dq/dz}
The thermodynamic limit $dq/dz$ of the quasienergy differential $dq_N/dz$ is given by
\begin{align}
	\frac{dq}{dz}&=2R_0(z)+\frac{2z^2}{R_0(z)}-\frac{i}{R_0(z)}\int_{\Gamma}\frac{v(\z)R_0(\z)d\z}{\z-z},\label{eq: quasi-eng}
\end{align}
where $\G=\cup_{j=1}^nc_{j,n}$ and $R_0=\sqrt{z^2-1}\sim z$ as $z\ra\infty.$
\ep
\begin{proof}
	To derive the thermodynamic limit for the quasienergy differentials, we need to modify the $g$-function set-up accordingly. In fact, similarly to \eqref{gfct for u}, we have
	\begin{align}
		g_{t+}(z)+g_{t-}(z)=2z^2+\tilde{W_j},\quad z\in \gamma_{j,N},
	\end{align}
	where all other requirements are exactly the same as for deriving the thermodynamic limit of quasimomentum and the constants $\tilde W_j$ are determined by the requirement that $g_t$ is analytic at $z=\infty$. Repeating the same argument of Theorem 3.18 in \cite{TW22}, taking the thermodynamic limit, we obtain
	
	\begin{align}
		\frac{2g_t}{R_0}=\sum_{m=0}\frac{ H_m}{z^{m+1}},
	\end{align}
	where \begin{align}
		H_m=-\int_{\Gamma}v(\xi)\int_{\xi}^1\frac{\mu^m d\mu}{R_0(\mu)}d\xi-\frac{1}{2\pi i}\oint_{A_0}\frac{2\xi^{m+2}d\xi}{R_0(\xi)},
	\end{align}
$\G=\sum_{j=1}^n c_{j,n}$ is the spectral support and $A_0$ is negative oriented large circle enclosing interval $[-1,1]$. 
	Thus, we have
	\begin{align}
		\frac{2g_t}{R_0}&=-\frac{i}{R_0(z)}\int_\Gamma v(\xi)\log\frac{R_0(\xi)R_0(z)+z\xi-\delta_0^2}{\xi-z}d\xi+\frac{1}{2\pi i}\oint_{A_0}\frac{2\xi^2}{(\xi-z)R_0(\xi)}d\xi\\
		&=-\frac{i}{R_0(z)}\int_\Gamma v(\xi)\log\frac{R_0(\xi)R_0(z)+z\xi-\delta_0^2}{\xi-z}d\xi+\frac{2z^2}{R_0(z)}-2z,
	\end{align}
	where the last two terms come from the residue theorem.
	It follows that
	\begin{align}
		2g_{tz}=-\frac{i}{R_0(z)}\int_{\Gamma}\frac{v(\xi)R_0(\xi)|d\xi|}{\xi-z}+4z-2R_0(z)-\frac{2z^2}{R_0(z)}.
	\end{align}
	Since $dq/dz=4z-2g_{tz}$, we have
		\begin{align}
			\frac{dq}{dz}=\frac{dq}{dz}&=2R_0(z)+\frac{2z^2}{R_0(z)}+\frac{i}{R_0(z)}\int_{\Gamma}\frac{v(\z)R_0(\z)d\z}{\z-z}.
		\end{align}
\end{proof}

Let us now consider  $\frac{dp}{dz}$ and $\frac{dq}{dz}$ for defNLS soliton condensates described in Theorem \ref{thm condensate}, where 
$u,v$ are proportional to the densities  of  the quasimomentum and quasienergy meromorphic differentials on the Riemann surface $\Rscr$. In particular,
  according to Remark \eqref{rem-dp-infty},
$u(z)=\frac{idp_\infty}{\pi dz}$.
 Substitute $u(z)$, see \eqref{eq: DOS+DOF}, into \eqref{quasi-bre}. Then \eqref{quasi-bre} becomes
\begin{align}\label{step1}
-\frac i{R_0(z)}\int_\G 
\frac{u(\z)R_0(\z)}{\z-z} |d\z|=
\frac 1{2\pi iR_0(z)}\oint_C \frac{\pi P(\z)d\z}{R(\z)(\z-z)} =
\frac{dp_\infty}{dz}(z)-\frac z{R_0(z)},
\end{align} 
where $C$ is a negatively oriented contour containing $\G$ but not containing $z$. Equation \eqref{step1} together with \eqref{quasi-bre} implies
\be\label{dpinfdp}
\frac{dp_\infty}{dz}=\frac{dp}{dz}.
\ee 
Thus, we obtain the following theorem.
\bt\label{th-limdp}
The quasimomentum differential $\frac{dp_\infty}{dz}$ on the ``superband" Riemann surface $\Rscr$, see \eqref{dpinfdp}, coincides 
with $\frac{dp}{dz}$, given by 
\eqref{quasi-bre}.
Similar statement is true for the quasienergy differentials.
\et

\label{cor-coinc}
\bc All the corresponding average densities and fluxes of defNLS soliton condensates with the spectral support $\G$ and of defNLS finite gap solutions on $\Rscr$ defined by $\G$ (see Remark \ref{rem-dp-infty}) coincide.
\ec	

\section{Kurtosis for the genus 0 and genus 1 defNLS gas}
\label{sec-kurt}

In this section, we compute the kurtosis $	\kappa$ for the defNLS gas,  which is defined by
\begin{align}
	\kappa = \frac{\braket{|\psi|^4}}{\braket{|\psi|^2}^2},\label{kur def}
\end{align}
where $\psi$ is a defNLS soliton gas realization and the bracket stands for the ensemble average.
 It is well known (\cite{FFM}, \cite{ForLee}) that the averaged conserved quantities (densities and fluxes) for the multiphase (finite gap) solution to the integrable PDEs can be computed by expanding the quasimomentum and quasienergy meromorphic differentials on the underlying Riemann surface. The corresponding computations for the averaged conserved quantities of the fNLS soliton/breather gases are derived by the authors in \cite{TW22}. In fact, based on Proposition \ref{theo-ImN} and \ref{prop:dq/dz}, we can compute the averaged conserved quantities (densities and fluxes) of the defNLS gases by computing the coefficients of \eqref{quasi-bre} and \eqref{eq: quasi-eng} respectively, where $\G$ is the spectral support.

\bc
\label{cor: dp coef}
The thermodynamic limits $dp/dz$ and $dq/dz$ of the  densities of the quasimomentum and quasienergy differentials respectively admit the following expansions as $z\ra\infty$:
\begin{align}
	\frac{dp}{dz}&=1-\sum_{m=1}^\infty\frac{I_m}{z^{m+1}},\\
	\frac{dq}{dz}&=1-\sum_{m=1}^\infty\frac{J_m}{z^{m+1}},
\end{align}
where
\begin{align}
	I_m&=\int_{\G} u(\z)\le[\frac{\z^m}{R_0(\z)}\ri]_+\sqrt{1-\z^2}d\z+md_m,\label{eq: Im} \\
	J_m&=\int_{\G} v(\z)\le[\frac{\z^m}{R_0(\z)}\ri]_+\sqrt{1-\z^2}d\z+2md_{m+1},\label{eq: Jm}
\end{align}
and $d_m$ is given by equation \eqref{dm-int}.
\ec

\begin{proof}
	According Proposition \ref{theo-ImN}, the thermodynamic limit of the defNLS quasimomentum differential $\frac{dp_N}{dz}$ is given by equation \eqref{quasi-bre}. Next, we expand \eqref{quasi-bre} at $z=\infty$. Recall that
	\[R_0(z)=\sum_{k=-1}^\infty d_kz^{-k},\]where $d_k$ is given by \eqref{dm-int} and $d_{-1}=1$. We also introduce $\tilde{d}_k$ so the reciprocal of $R_0(z)$ admits the following expansion at $\infty$: $1/R_0(z)=\sum_{k=1}^\infty \tilde d_kz^{-k}$. Since $R'_0(z)=z/R_0(z)$, we have $\tilde d_k=(2-k)d_{k-2},k\geq 1$. Moreover, the coefficient of $z^{-(m+1)}$ of $z/R_0(z)$ is $md_m$. 
	
	Expanding the integral term in \eqref{quasi-bre} we have
	\begin{align*}
		-\frac i{R_0(z)}\int_\G 
		\frac{u(\z)R_0(\z)}{\z-z} d\z&=i\int_{\G}u(\xi)\le(\sum_{k=1}^\infty \tilde d_kz^{-k}\ri)R_0(\z)\sum_{k=0}^\infty\z^{k}z^{-(k+1)}d\z\\
		&=\sum_{m=1}^\infty z^{-(m+1)}\int_{\G} iu(\z)\le(\sum_{k=1}^m\tilde d_k\z^{m-k}\ri)R_0(\z)d\z\\
		&=-\sum_{m=1}^\infty z^{-(m+1)}\int_{\G} u(\z)\le[\frac{\z^m}{R_0(\z)}\ri]_+\sqrt{1-\z^2}d\z
	\end{align*}
	
	Extracting the coefficient of $z^{-(m+1)}$, we then get 
	\begin{align}
		I_m = \int_{\G} u(\z)\le[\frac{\z^m}{R_0(\z)}\ri]_+\sqrt{1-\z^2}d\z+md_m.
	\end{align}
	
	Similarly, one can get the coefficients for the quasienergy differential $\frac{dq}{dz}$ by expanding \eqref{eq: quasi-eng}. In fact, we have
	\begin{align*}
		2R_0(z)+2z^2/R_0(z)=2R_0(z)+2zR_0(z)=\sum_{k=-1}^\infty (2d_k-2kd_k)z^{-k}.
	\end{align*}
	Thus, the coefficient of $z^{-(m+1)}$ is $-2md_{m+1}$. The coefficients extracting from the integral term of  \eqref{eq: quasi-eng} can be obtained similarly as for \eqref{eq: Im} by replacing $u$ with $v$. 
\end{proof}

Next, we compute the averaged densities and fluxes for the defNLS gases. The densities and fluxes for the defNLS equation can be derived using so-called quadratic eigenfunction method (see \cite{ForLee}). To compute the kurtosis, it suffices to use just the first few densities and fluxes. They are
\begin{align}
	f_1&=|\psi|^2,\\
	f_3&=|\psi|^4+|\psi_x|^2,\\
	g_2&=|\psi|^4+2|\psi_x|^2.
\end{align}
Based on Corollary \ref{cor-coinc} and the same normalization as for the fNLS circular gas in \cite{TW2023}, we get the following formulae for computing the averaged densities and fluxes for the defNLS gas:

\begin{align}
	I_1&=-\frac{1}{2}\braket{f_1}:=-\frac{1}{2}\braket{|\psi|^2},\label{df1}\\
	I_3&=-\frac{3}{8}\braket{f_3}:=-\frac{3}{8}\braket{|\psi|^4+|\psi_x|^2},\label{df2}\\
	J_2&=-\frac{1}{2}\braket{g_2}:=-\frac{1}{2}\braket{|\psi|^4+2|\psi_x|^2}.\label{df3}
\end{align}
Applying Corollary \ref{cor: dp coef} we obtain

\begin{align}
	I_1&=-1/2+\int_\G u(\z)\sqrt{1-\z^2}d\z,\label{I1}\\
	I_3&=-3/8+\int_\G (\z^2+1/2)u(\z)\sqrt{1-\z^2}d\z,\label{I3}\\
	J_2&=-1/2+\int_{\G}\z v(\z)\sqrt{1-\z^2}d\z, \label{J2}
\end{align}
where $u,v$ are the DOS and DOF for the defNLS gas, given by Theorem \ref{thm condensate}. Using equations \eqref{df1} ,\eqref{df2} and  \eqref{df3} one obtains
\begin{align}
	\braket{|\psi|^2}=2I_1=1-2\int_\G u(\z)\sqrt{1-\z^2}d\z,
\end{align}
and
\begin{align}
	\braket{|\psi|^4}&=-\frac{16}{3}I_3+2J_2\nonumber\\
	&=1-\frac{16}{3}\int_\G \le(\z^2+\frac{1}{2}\ri)u(\z)\sqrt{1-\z^2}d\z+2\int_{\G}\z v(\z)\sqrt{1-\z^2}d\z.
\end{align}

By the definition of kurtosis \eqref{kur def}, we have the following formula for computing the kurtosis of the defNLS condensate:
\begin{align}\label{kurt}
	\kappa&=\frac{\frac{1}{2}J_2-\frac{4}{3}I_3}{I_1^2}\nonumber\\
	&=\frac{1-\frac{16}{3}\int_\G \le(\z^2+\frac{1}{2}\ri)u(\z)\sqrt{1-\z^2}d\z+2\int_{\G}\z v(\z)\sqrt{1-\z^2}d\z}{(1-2\int_\G u(\z)\sqrt{1-\z^2}d\z)^2}.
\end{align}

We summarize the above computation as the following lemma.
\bl
\label{lem:kur}
For any defNLS condensate with spectral support $\G$, the kurtosis can be computed by the formula \eqref{kurt}, which admits another representation:
\begin{align}
	\kappa = \frac{1-\int_\G\le(\frac{16}{3}(\z^2+\frac{1}{2})P(\z)-2\z Q(\z)\ri)\frac{d\z}{R(\z)}}{\le(1+\int_\G2P(\z)\frac{d\z}{R(\z)}\ri)^2},
\end{align}
where polynomials $P,Q$ are given in Theorem \ref{thm condensate}.
\el

\begin{proof}
	Using the DOS($u$)/DOF($v$) representations (see equation \eqref{eq: DOS+DOF}) as given in Theorem\ref{thm condensate}, plugging in formula \eqref{kurt}, and after simplification, we obtain the kurtosis formula as stated.
\end{proof}

\br\label{rm:cre21}
The averaged conserved quantity (see equation \eqref{I1}) was also previously presented in \cite{CER21}, see equation (28) there. In fact, following the argument in \cite{CER21} (section III), one can derive the formulae \eqref{I3} and \eqref{J2} as well. The idea is to use the dark soliton solution (see \cite{ZS1973}) of the defNLS equation, which is given by 
\begin{align}
	\psi(x,t;z)=\frac{(z+i\sqrt{1-z^2})^2+A}{1+A}{e^{-2it}},\quad A=e^{2\sqrt{1-z^2}(x-2z t)}.
\end{align}
One can rewrite the solution in the form of 
\[\psi(x,t;z)=\sqrt{\rho}e^{iS(x,t)},\]
where
\begin{align}
	\rho(x,t)&:=|\psi(x,t;z)|^2=1-(1-z^2)\sech^2(\sqrt{1-z^2}(x-2zt)),\\
	S(x,t)&:=\text{Arg}(\psi)=\arctan\frac{2z\sqrt{1-z^2}}{A-1+2z^2},
\end{align}
which coincide (up to a factor $2$ in the linear dispersion relation, due to the different settings for the defocusing NLS equation) with equation (10) in \cite{CER21}, where $u_s^{\pm}$ in \cite{CER21} is just the $x$-derivative of the phase, $\partial_xS$.
Then, following \cite{CER21}, we can compute the ensemble average of the densities and fluxes for the defNLS soliton gas as follows:
\begin{align}
	I_3&=-\frac{3}{8}\braket{f_3[\psi(x,t;z)]}\\
	&=-\frac{3}{8}\le(\int_\G u(z)\le(\int_\R (f_3[\psi(x,t;z)]-1)dx\ri)dz+1\ri)\\
	&=-\frac{3}{8}\le(\int_\G u(z)\le(\int_\R (f_3[\psi(x,0;z)]-1)dx\ri)dz+1\ri).
\end{align}
The last step follows from the fact integral of $\int_\R(f_3-1)dx$ is invariant in $t$. A direct computation of the last integral gives \eqref{I3}. Similarly, one can get equation \eqref{J2} by computing the following integral:
\begin{align}
	J_2=-\frac{1}{2}\le(\int_\G v(z)\le(\int_{\R}(g_2[\psi(0,t;z)]-1)dt\ri)dz+1\ri)
\end{align}
\er

\subsection{Kurtosis for genus 1 and 0 defNLS condensates}
In this subsection, we restrict the discussion to the genus one and genus zero condensate cases, where the explicit expressions for $u,v$ were calculated in Section \ref{sec-condens}. The theorem below gives the   kurtosis formula for general
genus one defNLS condensate.

\bt
\label{genus1kurt}
For a general genus one defNLS condensate(i.e for each $\{a_j\}_{j=1}^4$ such that $-1\leq a_1<a_2<a_3<a_4\leq 1$), 
the corresponding kurtosis is given by the following expression:
\begin{align}
	\kappa=\frac{A\mu+B}{3(C\mu+D)^2},\label{g1 kurt formula}
\end{align}
where
\begin{align*}
	A&=-8 \left(-a_{4}+a_{2}\right) \left(-a_{3}+a_{1}\right)\le(4\sum_{j=1}^4a_j^2-(\sum_{j=1}^4a_j)^2 \right),\\
	B&=3 a_{1}^{4}+\left(12 a_{2}-12 a_{3}-12 a_{4}\right) a_{1}^{3}\\
	&+(2 a_{2}^{2}+\left(-20 a_{3}-20 a_{4}\right) a_{2}+18 a_{3}^{2}+20 a_{4} a_{3}+18 a_{4}^{2})a_1^2\\
	&+(12 a_{2}^{3}-4(a_3+a_4)(5 a_{2}^{2}-5 a_{3} a_{2}-5 a_{4} a_{2}+3 a_{3}^{2}+2 a_{4} a_{3}+3 a_{4}^{2}))a_1\\
	&+3 a_{2}^{4}+\left(-12 a_{3}-12 a_{4}\right) a_{2}^{3}+\left(18 a_{3}^{2}+20 a_{4} a_{3}+18 a_{4}^{2}\right) a_{2}^{2}\\
	&+\left(-12 a_{3}^{3}-20 a_{3}^{2} a_{4}-20 a_{3} a_{4}^{2}-12 a_{4}^{3}\right) a_{2}\\
	&+3 a_{3}^{4}+12 a_{3}^{3} a_{4}+2 a_{3}^{2} a_{4}^{2}+12 a_{3} a_{4}^{3}+3 a_{4}^{4}\\
	C&=4 \left(-a_{4}+a_{2}\right) \left(-a_{3}+a_{1}\right),\qquad
	D=-(a_1+a_2-a_3-a_4)^2,\\
	\mu&=\frac{E(m)}{K(m)}, \qquad
	m=\frac{\left(a_{3}-a_{4}\right) \left(-a_{2}+a_{1}\right)}{\left(a_{2}-a_{4}\right) \left(-a_{3}+a_{1}\right)}.
\end{align*}
\et
\begin{proof}
	Plugin the genus one DOS/DOF, which are given in Theorem \ref{thm condensate}, into the kurtosis formula \eqref{kurt}, direct computation gives the result. 
\end{proof}

\bc
\label{kurt: g0}
For a general genus zero defNLS condensate  the kurtosis is always 1.
\ec
\begin{proof}
	Using the DOS/DOF of the genus zero condensate (see Corollary \ref{g0-master case}), by the residue computations, we have
	\begin{align}
		\braket{|\psi|^2}=\frac{(a_1-a_2)^2}{4},\label{psi2}\\
		\braket{|\psi|^4}=\frac{(a_1-a_2)^4}{16}.\label{psi4}
	\end{align}
	Then it follows that  the kurtosis is 1 for any $-1<a_1<a_2<1$.
\end{proof}

\br
Using \eqref{df2} and equation \eqref{psi4}, we obtain, for genus zero condensate, we have
\begin{align}
	\braket{|\psi_x|^2}=\frac{(a_1^2-a_2^2)^2}{4}.\label{eq: psi_x}
\end{align}
\er
\br
In the vacuum case, the support of the realization of the dark soliton fill the whole interval $[-1,1]$ and we have $u=1/\pi$ and $v=4z/\pi$. By taking the limit $a_1\ra a_2$ in \eqref{psi2} and \eqref{psi4}, we obtain that	
\begin{align}
	\braket{|\psi|^4}=0,\quad \braket{|\psi|^2}=0.
\end{align}

\er

	Corollary \ref{kurt: g0} shows that,
	similarly to KdV soliton condensates (\cite{CERT}),
	the genus zero 
	soliton condensate is almost surely given by a constant amplitude solution 
	to the defNLS since the variance $\sqrt{\braket{|\psi|^4}  -	\braket{|\psi|^2}^2}=0$.  Moreover, in the case  $a_1=-a_2$, it follows from  \eqref{eq: psi_x}  that the phase $\arg \psi$ is almost surely a constant.  
	We now analyze the general case $a_1\neq -a_2$.

The  Madelung transformation reduces the NLS \eqref{NLS} to the hydrodynamic type system 
\be\label{madelung}
\hf	\rho_t+ (\r\m)_x=0, \qquad
\hf(\r\m)_t+(\r\m^2-s\hf\r^2)_x=\frac 14[\r(\ln \r)_{xx}]_x,
\ee
where $\r=|\psi|^2$, $\m=(\arg \Psi)_x$ and $s=-1$ for defNLS. If $\psi$ is a generic realization of  defNLS genus zero soliton condensate then $\r_t=0$ and from the first \eqref{madelung} equation we obtain $\arg\psi=k(t)x+c(t)$ for some functions $k(t),c(t)$ to be determined.
Substituting $\m=k(t)$ into the second \eqref{madelung} equation 	we see that $k$ is independent of $t$. Thus,
\be\label{gen-0-sol}
\psi(x,t)=\frac{b-a}2e^{i(kx+c(t))}.
\ee	
Now, using \eqref{eq: psi_x}, we  find $k=(b+a)$. Substituting \eqref{gen-0-sol} to \eqref{NLS}, we find $c(t)=\o t$ with some constant $\o$. Thus, the realization we consider almost surely has the form 
\be 
\psi(x,t)=\hf(b-a)e^{i[(b+a)x+\o t]},
\ee
which is also the form of a plane wave solution defined by a spectral band $[a,b]$
(up to a constant phase).
Thus, we extended the observation of \cite{CERT} that {\it a genus zero KdV condensate almost surely coincides with the corresponding  genus zero KdV solution for defNLS genus zero condensates.}

According to Theorem \ref{th-limdp} and its Corollary \ref{cor-coinc}, the averaged conserved quantities for the genus one defNLS condensate coincides with  the averaged conserved quantities for a genus one {finite gap solution} of the defNLS equation constructed from the same Riemann surface. Thus, to compute the kurtosis for the defNLS condensate is equivalent to compute the kurtosis for
a genus one   {finite gap solution} of the defNLS equation. In fact, to compute the kurtosis, we only need to know the {modulus} of the genus one solution to the defNLS equation. Given the same Riemann surface as for the defNLS condensate, it is well known(see \cite{KamBook}, chapter 5, Section 2) {that}
\begin{align}
	|\psi(x,t)|^2=\frac{1}{4}(a_1-a_2-a_3+a_4)^2+(a_2-a_1)(a_4-a_3)\text{sn}^2\le(\Lambda(x,t),m\ri),
\end{align}
where
$
	\Lambda(x,t)=\sqrt{(a_4-a_2)(a_3-a_1)}\le(x-t\sum_{j=1}^4a_j\ri),
$
where  $m=\frac{\left(a_{3}-a_{4}\right) \left(-a_{2}+a_{1}\right)}{\left(a_{2}-a_{4}\right) \left(-a_{3}+a_{1}\right)}$ is {given by}
\eqref{g1 kurt formula}.
Then by definition of the kurtosis we get :
\begin{align}
	\kappa =\frac{\frac{1}{T}\int_{0}^T|\psi(x,t)|^4dx}{\le(\frac{1}{T}\int_0^T|\psi(x,t)|^2dx\ri)^2}= \frac{T\int_{0}^T|\psi(x,0)|^4dx}{\le(\int_0^T|\psi(x,0)|^2dx\ri)^2},\label{eq:new k rep}
\end{align}
where $T=2K(m)/\sqrt{(a_4-a_2)(a_3-a_1)}$ is the period of $|\psi(x,0)|$.
 Note that the second equality follows from the fact that varying $t$ only shifts the genus-one finite-gap solution but without changing its spatial average.   Using the new representation, we can show that  the kurtosis for the genus one condensate is always bounded by 3/2 from above and bounded (trivially) by 1 from below.

\bt
For general genus one defNLS condensate, the kurtosis satisfies the following inequality:
\[1\leq \kappa\leq 3/2.\]
\et
\begin{proof}
	The lower bound is a consequence of Cauchy-Schwartz inequality. Next we prove the upper bound. 
	
	Based on the new representation (\eqref{eq:new k rep}) of the kurtosis  in terms of integrals of Jacobi theta functions, it is convenient to introduce some new notations:
	\begin{align}
		b=\frac{1}{4}(a_1-a_2-a_3+a_4)^2,\quad A=(a_2-a_1)(a_4-a_3).
	\end{align}
	Then equation \eqref{eq:new k rep} becomes
	\begin{align}
		\kappa=\frac{T\int_0^T\le[b+A\,\text{sn}^2(\Lambda(x,0),m)\ri]^2dx}{\le(\int_0^T [b+A\,\text{sn}^2(\Lambda(x,0),m)]dx\ri)^2},
	\end{align}
	where using the new notation, $\Lambda(x,0)=\sqrt{A/m}x$.
	First, assuming $A,b$ and $m$ are independent variables and fixing $m,A$, we find {after simple calculations}
	\begin{align*}
		\partial_b\kappa&=2T\frac{\le(\int_0^T\le(A\,\text{sn}^2(\Lambda(x,0),m)+b\ri)dx\ri)^2-T\int_0^T\le(A\text{sn}^2(\Lambda(x,0),m)+b\ri)^2dx}{\le(\int_0^T\le((A\text{sn}^2(\Lambda(x,0),m)+b\ri)dx)\ri)^3}\leq 0.
	\end{align*}
	where with used that, according to   the  Cauchy-Schwartz inequality,
	\begin{align*}
		&\le(\int_0^T\le(A\,\text{sn}^2(\Lambda(x,0),m)+b\ri)dx\ri)^2\leq
		T\le(\int_0^T\le(A\,\text{sn}^2(\Lambda(x,0),m)+b\ri)^2dx\ri).
	\end{align*}

	Thus, we have $\kappa\leq \kappa|_{b=0}.$
	When $b=0$, the kurtosis becomes
	\begin{align*}
		\kappa(m):=\kappa|_{b=0}&=\frac{T\int_0^T\text{sn}^4(\sqrt{A/m}x,m)dx}{(\int_0^T\text{sn}^2(\sqrt{A/m}x,m)dx)^2}	
		=\frac{2K(m)\int_0^{2K(m)}\text{sn}^4(y,m)dy}{(\int_0^{2K(m)}\text{sn}^2(y,m)dy)^2}.
	\end{align*}
Using formulae (310.02) and (310.04) in \cite{BFbook}, the expression for the kurtosis can be further simplified to 
\begin{align}
	\kappa(m)=\frac{2+m-2(1+m)\mu}{3(\mu-1)^2},	
\end{align}
where $\m$ is given in \eqref{g1 kurt formula}.

	To show $\kappa(m)\leq 3/2$, it is sufficient to show
	\begin{align}
		F(\mu)=9\mu^2+2(2m-7)\mu+5-2m \geq 0,\forall m\in[0,1].
	\end{align}
	The discriminant of $F(\mu)$ (regard $\mu$ as independent variable) is given by $ 8(2m-1)(m-2)$. Since $m\leq 1$, the discriminant of $F$ is negative for $m>1/2$. And since the leading coefficient of $F$ is positive, we then have shown that for $m\geq 1/2$, $F\geq 0$, which in turn implies $\kappa(m)\leq 3/2$ for $m\geq 1/2$. 
	
	Next, we show that $\kappa(m)\leq 3/2$ for $m\in [0,1/2]$. Using Theorem  \ref{low bd mu}, we have
	 \be\mu(m):=\frac{E}{K}\geq   1-m/2-m^2/16-\frac{7}{100} m^3.\label{mu low bound}\ee

 To prove that $F(\mu)>0$,it is sufficient to show the lower bound \eqref{mu low bound} is larger than the largest root of $F$, which is $\frac{1}{9}(7-2m+\sqrt{2(2m-1)(m-2)})$. Since
	\begin{align*}
		&\le[9( 1-m/2-m^2/16-\frac{7}{100}m^3)-(7-2m)\ri]^2-2(2m-1)(m-2)\\
		&=2(1-2 m) +\frac{117}{400} m^{3}+\frac{4437}{1280}  m^{4}+\frac{567}{800} m^{5}+\frac{3969}{10000} m^{6}>0, \,\, \forall m\in [0,1/2],
	\end{align*}
	we have
	$$
	\mu(m)\geq 1-m/2-m^2/16-\frac{7}{100}m^3>\frac{1}{9}(7-2m+\sqrt{2(2m-1)(m-2)}),
	$$
	which then implies $F(\mu)>0$ whenever $m\in[0,1/2]$.
	
	Thus, for all $m\in[0,1]$, we have $F\geq 0$, which in turn implies $\kappa(m)\leq 3/2$ for any $m\in [0,1]$.
\end{proof}

\section{Diluted defNLS condensate}
\label{sec-diluted}
Consider now the diluted genus 0 defNLS condensate with $u(z)=\frac \a\pi$, where $\a\in(0,1)$. It follows then from \eqref{1stNDR} that
\be\label{dil-sig}
\s(z)=\frac{\pi(1-\a)}\a |R_0(z)|.
\ee
We use \eqref{quasi-bre} and \eqref{Hil-sq} to calculate
\be\label{dil-dp}
\frac{	dp}{dz}=\a+(1-\a)\frac z{R_0(z)}	=1+(1-\a)\le(\frac 1{2z^2}+\frac 3{8z^4}+\dots \ri),
\ee
which expresses the average conserved densities through the binomial coefficients of the expansion of  $\frac 1{\sqrt{1-z^{-2}}}$ at $z=\infty$.
In particular,
\be\label{I1,3}
I_1=\frac {\a-1}2,\qquad I_3=\frac {3(\a-1)}8,
\ee
etc., where 
\be\label{dp-dil}
\frac{dp}{dz}=1-\sum_{m=1}^\infty\frac{I_m}{z^{m+1}}.
\ee
Given that $\langle|\psi|^2\rangle=-2I_1$ , we obtain 
\be\label{1st-mom-dil}
\langle|\psi|^2\rangle=1-\a,
\ee
i.e., we have the vacuum $\langle|\psi|^2\rangle=0$ in the case of the condensate $\a=1$.

We are now trying to solve the 2nd NDR \eqref{FTH-v} with $\s$ given by \eqref{dil-sig} by $v(z)=bz$. Then
\be\label{dif-term-dil}
\sqrt{1-z^2} \frac d{dz}[\s(z)v(z)]=\frac{b\pi(1-\a)}\a [1-2z^2],
\ee
so that 
\be\label{FTH-v-dil}
\frac{b\pi}2(2z^2-1)=\le(2+\frac{b\pi(1-\a)}\a \ri)(2z^2-1) \quad \Leftrightarrow \quad b=\frac{4\a}{\pi(2-\a)}.
\ee

Thus, 
\be
v(z)=\frac{4\a z}{\pi(2-\a)}
\quad\text{and}\quad s(z)=\frac{4z}{2-\a},
\ee
which shows that the effective velosity $s(z)$ varies
from velocity of  free dark solitons $s(z)=2z$ in the diluted gas limit $\a=0$ to the double of this speed $s(z)=4z$ in the condensate (vacuum) limit $\a=1$.

To calculate the kurtosis $\kappa$, see \eqref{kurt}, of the diluted condensate, we use \eqref{J2} with $v(z)=bz$ that yields
\be
\int_{-1}^1\z^2\sqrt{1-\z^2}d\z=\frac{-i}2\oint\z^3\le(1-\frac 1{2\z^2}-\frac 1{8\z^4}-\dots\ri)d\z=\frac \pi 8,
\ee
so that 
\be\label{J2-dil}
J_2=-\hf+\frac{4\a}{\pi(\a-2)}\cdot \frac \pi 8=\frac {\a-1}{2-\a}.
\ee
Now, according to \eqref{kurt},
\be\label{kurt-dil}
\kappa=\frac 2{2-\a},
\ee
which shows that the kurtosis varies from $\kappa=1$ in the plane wave limit $\a=0$ and $\kappa=2$ in the condensate (vacuum) limit $\a=1$, the latter being consistent with the Gaussian statistics.

 	\appendix
 \section{An inequality on complete elliptic integrals}
 In this appendix, we give a proof on a lower bound of $\mu(m)=E(m)/K(m)$, where $E,K$ are the complete elliptic integrals of the first and second kind respectively and $m$ is the elliptic modular parameter. 
 \bt
 \label{low bd mu} For $m\in [0,1/2]$, $\mu$ satisfies the following inequality:
 \begin{align}
 	\mu(m)>1-m/2-m^2/16-\frac{7}{100}m^3. \label{eq: low}
 \end{align}
 \et

 Before we prove the theorem, we first prove some properties of the function $\mu$. Recall the definitions of the complete elliptic integrals of the first and the second kind:
 \begin{align}
 	E(m)=\int_0^{\pi/2}\sqrt{1-m\sin^2y}dy,\quad K(m)=\int_0^{\pi/2}\frac{1}{\sqrt{1-m\sin^2y}}dy,\label{eq: def EK}
 \end{align}
 where $m\in[0,1]$. It is well-known (see for example 
 \cite{BFbook}) that $E(k),K(m)$ admit convergent power series expansions near $m=0$ with radius of convergence 1 and both are positive
  for $m\in [0,1)$.

 \bl
 For any $m\in [0,1)$, $\mu$ satisfies the following Riccati equation:
 \begin{align}
 	\frac{d\mu}{dm}=\frac{\mu^2+2(m-1)\mu-(m-1)}{2m(m-1)}.\label{eq: diff rel}
 \end{align}
 \el
 \begin{proof}
 	It is well-known (see 
 	\cite{BFbook}, formulas (710.00) and (710.02)) that for $m\in [0,1)$
 	\begin{align}
 		E'(m)=\frac{E-K}{2m},\quad K'(m)=\frac{E-(1-m)K}{2m(1-m)},
 	\end{align}
 	then direct computation leads to the following Riccati equation:
 	\begin{align}
 		\mu'(m)=\frac{E'K-EK'}{K^2}=\frac{\mu^2+2(m-1)\mu-(m-1)}{2m(m-1)}.
 	\end{align}
 \end{proof}
 
 Since $K(0)=\pi/2>0$, the quotient $\mu=E/K$ admits a convergent power series expansion near $m=0$ with positive radius of convergence.
 Let  
 \begin{align}
 	\mu(m)=\sum_{j=0}^\infty a_jm^j.\label{eq: mu expan}
 \end{align}
It is obvious that $a_0=1$.
 We first prove  that the radius of convergence of the power series is at least 1/2. Then we show that all the coefficients $a_j,j\geq 1$ are negative.
 \bl
 The radius of convergence of the power series expansion \eqref{eq: mu expan} is at least $\frac{1}{2}$.
 \el
 \begin{proof}
 	Since the radius of convergence of $E(m)$ is 1 and $\mu$ is the product of $E$ and the reciprocal of $K$, it is sufficient to show the radius of convergence of $1/K$ is at least 1/2. Denote the power series expansion of $\frac{2}{\pi}K$ by $\sum_{j=0}^\infty k_jm^j$ for $m\in (-1,1)$ and from the formula (900.00) in \cite{BFbook} we know $k_0=1$ and
 	 $k_j>0,j>0$. Denote the power series expansion of the reciprocal of $\frac{2}{\pi}K$ near 0 by $\sum_{j=0}^\infty \tilde k_jm^j$, then the coefficients $\tilde k_j$ are determined recursively:
 	\begin{align}\label{rec}
 		\tilde{k}_0=1,\tilde k_n=-\sum_{j=1}^nk_j\tilde k_{n-j},\quad n\geq 1.
 	\end{align} 
 	It follows from \eqref{eq: def EK} that $K$ is an increasing function of $m$. Then one can show that for any $m\in[0,\hf] $:
 	\begin{align}\label{est}
 		\sum_{j=1}^\infty k_jm^j=\frac{2}{\pi}K(m)-1\leq \frac{2}{\pi}K(1/2)-1<1,
 	\end{align}
 	where the last inequality follows from 
 	\begin{align*}
	 		\frac{2}{\pi}K\le(\frac{1}{2}\ri)-1&=\int_0^{\pi/2}\frac{2dy}{\pi\sqrt{1-\frac{1}{2}\sin^2y}}-1\\
	 		&<\int_0^{\pi/2}\frac{2dy}{\pi\sqrt{1-1/2}}-1=\sqrt{2}-1<1.
 	\end{align*}
 	Next we proof $|\tilde k_j|\leq 2^j$ by induction. For $j=0$, it is true that $\tilde{k}_0=1\leq 1$.   Suppose the estimate is true for $j\leq n-1$. Using 
 	\eqref{rec}, \eqref{est},
 	we obtain
 	\begin{align}
 		|\tilde k_n|\leq \sum_{j=1}^nk_j|\tilde k_{n-j}|\leq \sum_{j=1}^nk_j|2^{n-j}|<2^n\sum_{j=1}^\infty k_j(1/2)^j<2^n,
 	\end{align} 
 	and the claim follows.  	The proof is completed.
 \end{proof}

 \bl\label{lem pos}
 The coefficients $a_j,j\geq 1$ in the expansion \eqref{eq: mu expan} are negative.
 \el
 \begin{proof}
 	Pluging in the power series into the differential relation \eqref{eq: diff rel}, we obtain
 	\begin{align}
 		a_0=1,\quad a_1=-1/2,\quad -2ja_j=(4-2j)a_{j-1}+\sum_{k=1}^{j-1}a_ka_{j-k},\quad j\geq 2.\label{eq:rec rel}
 	\end{align}
 	Let $b_j=-a_j,j>0$, then the recursion relation \eqref{eq:rec rel} becomes
 	\begin{align}
 		2jb_j=(2j-4)b_{j-1}+\sum_{k=1}^{j-1}b_jb_{j-k},\quad j\geq 2.
 	\end{align}
 	We will show that $b_j>0$ by induction. First, we check $b_1=-a_1=1/2>0$. Suppose $b_k>0$ for $k=0,\cdots, j-1$, using the recursion relation, we have
 	\begin{align}
 		b_j\geq (1-2/j)b_{j-1}\geq \cdots \geq \frac{(j-2)!2!}{j!}b_2=\frac{1}{8}\le(\frac{1}{j-1}-\frac{1}{j}\ri)>0,\quad j\geq 2.
 	\end{align}
 	Thus, by induction, we have $b_j>0$ for all $j>0$, which implies that all $a_j, j>0$, are negative.
 \end{proof}

 \begin{proof}[Proof of Theorem \ref{low bd mu}]

 	Rewrite the power series expansion of $\mu$ in the following way:
 	\begin{align}
 		\mu(m)=1-m/2-m^2/16-(1/32+H(m))m^3,
 	\end{align}
 	where $H(m)=\sum_{j=4}^\infty (-a_j)m^{j-3}$. 
 	
 	Since all $a_j<0$, $H$ is an increasing function of $m$. We have, for $m\in [0,1/2]$,
 	\begin{align}
 		\mu(m)\geq 1-m/2-m^2/16-(1/32+H(1/2))m^3, \label{sharp inequality}
 	\end{align}
 	where $H(1/2)=187/32-8\mu(1/2)$. To prove the inequality \eqref{eq: low}, it suffices to check $1/32+H(1/2)<7/100$. Checking the value (see \cite{BFbook}, the first table in the appendix) of $E(m),K(m)$ at $m=1/2$, we find $E(1/2)>1.35$ and $K(1/2)<1.86$, hence $\mu(1/2)>135/186$, which then implies
 	\begin{align}
 		1/32+H(1/2)<1/32+187/32-8(135/186)=17/248<7/100,
 	\end{align}
 	and then the inequality \eqref{eq: low} follows.
 	
 \end{proof}

\end{document}